\documentclass[acmsmall]{acmart}

\AtBeginDocument{%
  \providecommand\BibTeX{{%
    \normalfont B\kern-0.5em{\scshape i\kern-0.25em b}\kern-0.8em\TeX}}}

\setcopyright{acmcopyright}
\acmJournal{PACMHCI}
\acmYear{2022} \acmVolume{6} \acmNumber{CSCW2} \acmArticle{362} \acmMonth{11} \acmPrice{15.00}\acmDOI{10.1145/3555087}

\usepackage{makecell}
\usepackage{siunitx} %align numbers by decimal point
\NewDocumentCommand{\anote}{}{\makebox[0pt][l]{$^*$}}
\usepackage{enumitem}   
\usepackage{comment}
\usepackage{multirow}
\usepackage{float}
\usepackage{tcolorbox}
\usepackage{quoting} %for quotations
\quotingsetup{vskip=0pt} %align a horizontal space before a quotation
\usepackage[colorinlistoftodos]{todonotes}
\usepackage{hyperref}

\usepackage{soul}
\definecolor{mediumblue}{rgb}{0.0, 0.0, 0.8}

\begin{document}

\title{Why, How and Where of Delays in Software Security Patch Management: An Empirical Investigation in the Healthcare Sector}
% \title{The Struggle Is Real: Exploring Why, How and Where of Delays in Software Security Patch Management}
% \title{The Struggle Is Real: Exploring Why, When and How to on the Delays in Software Security Patch Management}
% \title{The Struggle Is Real: Why, How, When and What to do about Delays Occur in Software Security Patch Management}
% \title{Delays in Software Security Patch Management: Understanding Why, How, When and What to do?}
% \title{The Real Struggle With Delays In Software Security Patch Management: Why, How, When and What to do?}
% \title{The Struggle Is Real: An Empirical Study of Delays in Software Security Patch Management}

\author{Nesara Dissanayake}
\affiliation{%
 \institution{CREST – Centre for Research on Engineering Software Technologies, The University of Adelaide}
 \country{Australia}}
\email{nesara.madugodasdissanayakege@adelaide.edu.au}

\author{Mansooreh Zahedi}
\affiliation{%
 \institution{The University of Melbourne}
 \country{Australia}}
\email{mansooreh.zahedi@unimelb.edu.au}

\author{Asangi Jayatilaka}
\affiliation{%
 \institution{CREST – Centre for Research on Engineering Software Technologies, The University of Adelaide}
 \country{Australia}}
\email{asangi.jayatilaka@adelaide.edu.au}

\author{Muhammad Ali Babar}
\affiliation{%
  \institution{CREST – Centre for Research on Engineering Software Technologies, The University of Adelaide}
  \country{Australia}}
\email{ali.babar@adelaide.edu.au}

\renewcommand{\shortauthors}{Dissanayake, et al.}

\begin{abstract}
  Numerous security attacks that resulted in devastating consequences can be traced back to a delay in applying a security patch. Despite the criticality of timely patch application, not much is known about why and how delays occur when applying security patches in practice, and how the delays can be mitigated. Based on longitudinal data collected from 132 delayed patching tasks over a period of four years and observations of patch meetings involving eight teams from two organisations in the healthcare domain, and using quantitative and qualitative data analysis approaches, we identify a set of reasons relating to technology, people and organisation as key explanations that cause delays in patching. Our findings also reveal that the most prominent cause of delays is attributable to coordination delays in the patch management process and a majority of delays occur during the patch deployment phase. Towards mitigating the delays, we describe a set of strategies employed by the studied practitioners. This research serves as the first step toward understanding the practical reasons for delays and possible mitigation strategies in vulnerability patch management. Our findings provide useful insights for practitioners to understand what and where improvement is needed in the patch management process and guide them towards taking timely actions against potential attacks. Also, our findings help researchers to invest effort into designing and developing computer-supported tools to better support a timely security patch management process.
\end{abstract}

\begin{CCSXML}
<ccs2012>
   <concept>
       <concept_id>10002978.10003022.10003023</concept_id>
       <concept_desc>Security and privacy~Software security engineering</concept_desc>
       <concept_significance>500</concept_significance>
       </concept>
   <concept>
       <concept_id>10011007.10011074.10011111.10011696</concept_id>
       <concept_desc>Software and its engineering~Maintaining software</concept_desc>
       <concept_significance>500</concept_significance>
       </concept>
   <concept>
       <concept_id>10002978.10003006.10011634</concept_id>
       <concept_desc>Security and privacy~Vulnerability management</concept_desc>
       <concept_significance>300</concept_significance>
       </concept>
 </ccs2012>
\end{CCSXML}

\ccsdesc[500]{Security and privacy~Software security engineering}
\ccsdesc[500]{Software and its engineering~Maintaining software}
\ccsdesc[300]{Security and privacy~Vulnerability management}

\keywords{patch management, security updates, delays, socio-technical research}

\maketitle

\section{Introduction}
\label{section:introduction}
% Remediation of the identified security vulnerabilities by applying patches on time is crucial to system security. Several security attacks that caused catastrophic consequences can be attributed to delays in patching security vulnerabilities for which a patch had been already existing for months. For example, the Equifax data breach that exposed sensitive data of 143 million people \cite{Equifaxwired, Equifaxbug}, and the recent cyberattack in a German hospital that resulted in human death from delayed treatment \cite{Germancyberattack}. Yet, the recent statistics \cite{Accenturereport2020} reveal that the situation has not improved demonstrating serious concerns over delayed patching in practice. As such, demonstrated criticality of timely patching has prompted increased importance of the efforts aimed at reducing delays in software security patch management, referred to as security patch management hereafter, a process consisting of detecting, retrieving, assessing, installing, and verifying security patches \cite{souppaya2013guide}. Despite the gravity of the process, security patch management remains a challenging endeavour due to its inherent technical and socio-technical interdependencies \cite{li2019keepers, tiefenau2020security, dissanayake2021grounded}. For example, security patch management is an inherently collaborative process requiring effective coordination to resolve the interdependencies of diverse stakeholders in dispersed locations with contrasting interests. 

Cyberattacks breaching corporate networks often result in catastrophic consequences ranging from exposure of sensitive and confidential data \cite{Equifaxwired, Equifaxbug} and betrayal of client trust to even human death \cite{Germancyberattack}. The most effective remediation of this problem is to apply security patches to the identified vulnerabilities through a process called software security patch management, referred to as security patch management hereafter, consisting of detecting, retrieving, assessing, installing, and verifying security patches \cite{souppaya2013guide}. Despite the gravity of the process, security patch management remains one of the most challenging endeavours due to the inherent technical and socio-technical interdependencies involved in the collaborative process of dealing with third-party vulnerabilities and vendor patches \cite{li2019keepers, tiefenau2020security, dissanayake2021grounded}. As a result, organisations struggle to apply timely patches often leaving myriad vulnerabilities open to exploits. Consequently, it has resulted in most security attacks targeting known vulnerabilities for which a patch existed but delayed application. Despite the demonstrated criticality of timely patching, the recent statistics \cite{Accenturereport2020} reveal that the situation has still not improved indicating serious concerns and the increased importance of the efforts aimed at reducing delays in security patch management in practice.

% While the previous studies have investigated the socio-technical aspects of security patch management, particularly, the challenges \cite{li2019keepers, tiefenau2020security} and the role of coordination in the process \cite{dissanayake2021grounded, nicastro2003security}, these studies have not exclusively focused on the causes of delays in applying security patches. A limited number of studies have attempted to optimise the process by focusing on synchronising the organisational patch cycle with the vendor \cite{nappa2015attack, cavusoglu2008security, cavusoglu2006economics}. Although the existing studies have focused on approaches to reduce delays in security patch management, to date, there has been no study that comprehensively explores why and how delays continue to happen when applying security patches. It adds to the demonstrated critical need for investigating the delays in patching, grounded in evidence from practice. Motivated by the need, we conducted a longitudinal study investigating the delays in security patch management. Our study was guided by the following key research questions (RQs):

While the previous studies have investigated the socio-technical aspects of security patch management, particularly, the process followed \cite{li2019keepers, tiefenau2020security} and the role of collaboration in the process \cite{nicastro2003security}, these studies have not exclusively focused on the delays in applying security patches. Further, another set of studies has attempted to optimise the patch management process by synchronising the organisational patch cycle with the vendor's patch release cycle \cite{nappa2015attack, cavusoglu2008security, cavusoglu2006economics}. Focused on the coordination aspect, our previous study \cite{dissanayake2021grounded} presented a grounded theory of the role of coordination in security patch management based on observations of 51 patch meetings between two case organisations over nine months. The theory explains the causes that define the need for coordination in the process (i.e., the socio-technical dependencies), constraints that can hinder effective coordination, the breakdowns resulting from ineffective coordination of the causes and constraints, and the mechanisms to manage the causes, constraints and breakdowns. Although previous studies have focused on approaches to reduce delays in security patch management and the effects of ineffective coordination on timely security patch management, to date, there has been no study that comprehensively explores why and how delays continue to happen when applying security patches. It adds to the demonstrated critical need for investigating the delays in patching, grounded in evidence from practice. Motivated by the need, we extended the longitudinal study of the two case organisations focusing on the delays in security patch management. The study findings are based on the analysis of the artefacts over four years following the Straussian Grounded Theory method for the data analysis. Our study was guided by the following key research questions (RQs):
 \begin{itemize}[label={}]
   \item \textbf{RQ1.} \textit{Why, how, and where do delays occur in software security patch management? }
% %   \todo{``How" answers questions like \textit{by what method?}, \textit{to what degree?}, \textit{in what condition?} and many more. Whereas, ``why" answers questions like \textit{for what purpose or reason?}}
% %   \begin{itemize}[label={}]
% %     \item \textbf{RQ1.1.} What causes delays when applying security patches?
% %     \item \textbf{RQ1.2.} How often do these causes of delays occur?
% %     \item \textbf{RQ1.3.} When do delays occur during the patch management process?
% %   \end{itemize}
   \item \textbf{RQ2.} \textit{How can the delays be mitigated?}
 \end{itemize} 

% It should be noted that we started the study with two broad RQs (RQ 1.1 and 2) while the other sub RQs emerged during the data analysis. As explained by previous studies \cite{stol2016grounded, urquhart2012grounded, rodriguez2020theory}, RQs can and do evolve during the study as new concepts and categories emerge when using the Grounded Theory data analysis procedures. 

Based on qualitative and quantitative analysis of the longitudinal data gathered from patch meeting minutes spanning over four years from October 2016 to May 2021 between two organisations in the healthcare domain, we attempt to answer these crucial overarching questions of delays in security patch management. The findings explain the causes of delays with a taxonomy comprising technology, people and organisation-related reasons and describe which reasons are more prominent based on their frequency distribution, and where the delays occur in the patch management process. This study also reports a classification of strategies applied in practice to mitigate the delays including when to apply them during the patch management process. To the best of our knowledge, this is the first study to provide a comprehensive understanding of the causes and strategies for delays in security patch management.

Grounded in descriptive evidence from practice, our research contributes to the state-of-the-art understanding of research and practice in several ways: (i) identifies a set of reasons for delays when applying security patches in practice; (ii) describes the most prominent reasons for delays with rationales explaining their variations; (iii) reports where a majority of delays occur in the patch management process presenting their distribution over the process phases; (iv) presents a collection of strategies employed in practice to mitigate the delays including when to apply them in the patch management process; (v) structures the understanding about delays in vulnerability patch management, drawing attention to a critical yet less explored phenomenon in the CSCW community; (vi) grounded in practical evidence, the findings lay a foundation for future researchers and tool designers to design and develop computer-supported solutions to reduce delays in patch application, and (vii) offers practical guidance for practitioners to identify what and where is improvement needed to mitigate patching delays and drive their decisions appropriately. 

\section{Background and Motivation}
\label{section:background}

\begin{figure*}[h]
  \centering
  \includegraphics[scale=0.8]{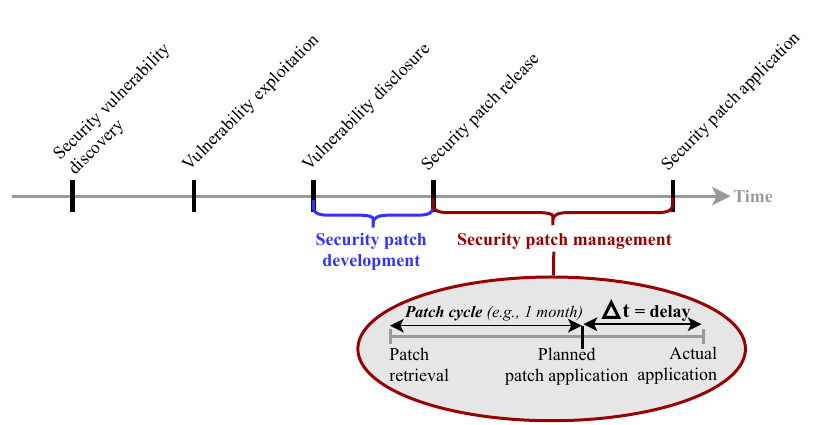}
  \caption{The focus of the study in the vulnerability timeline.}
  \Description{An overview of the vulnerability timeline showing a delay in software security patch management.}
  \label{fig:delayfocus}
\end{figure*}

Software security patch management is defined as \textit{``a multifaceted process of identifying, acquiring, testing, installing, and verifying security patches for software products and systems"} \cite{dissanayake2021software}. A security patch is an additional piece of code developed to address security vulnerabilities identified in software \cite{mell2005creating}. Following the discovery of a new vulnerability, a candidate security patch is developed and released by third-party vendors to prevent exploitation by malicious entities. For example, the Meltdown \cite{217478} and Spectre \cite{kocher2019spectre} patches released in 2018 by vendors such as Microsoft, Google, IBM and Apple were aimed at fixing two critical vulnerabilities in modern processors allowing malicious programs to gain unauthorised access to the software system. In security contexts, patch management represents a critical concern in achieving and maintaining the security of the managed software systems. This is because applying a security patch is considered the most effective mechanism to mitigate the identified vulnerabilities \cite{souppaya2013guide}. Similarly, applying security patches with minimum delays is instrumental in significantly reducing the risks of cyberattacks that exploit software vulnerabilities (see Figure \ref{fig:delayfocus}) \cite{souppaya2013guide}. Despite the importance of timely patch management, it remains one of the most challenging processes facing modern organisations. To guide the process, several guidelines such as the National Institute of Standards and Technology (NIST)’s Special Publication (SP) 800-40 \cite{nist2002guide, nist2005guide, souppaya2013guide} have been published over the years.

\begin{figure*}[h]
  \centering
  \includegraphics[scale=0.9]{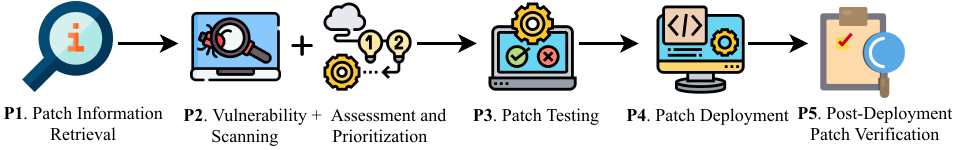}
  \caption{The five phases of the security patch management process.}
  \Description{An overview of the security patch management process.}
  \label{fig:pmprocess}
\end{figure*}

There have been several research efforts undertaken to explore the patch management process. Two recent studies \cite{li2019keepers, tiefenau2020security} have investigated the stages the practitioners proceed through patch management. Figure \ref{fig:pmprocess} shows an overview of the five main phases in the process. These phases represent the general workflow in a patch cycle (i.e., from the patch retrieval to planned patch application) depicted in Figure \ref{fig:delayfocus}. 

The patch information retrieval (P1) phase refers to learning about new patches and downloading them from vendor websites. Next, practitioners scan systems to identify the existing vulnerabilities, assess them based on their applicability to managed systems, and prioritise based on vulnerability severity and patch type when deciding to patch (P2). Once the decision to patch is finalised, they prepare to deploy the patches whereby testing the patches for accuracy (P3) and preparing machines by changing configurations, followed by the patch deployment (P4) phase in which the patches are installed and rebooted. As the final step, the patch deployment is verified and post-deployment issues are handled, if any (P5).    

In addition, prior research has invested effort in improving the patch management process through both technical and socio-technical aspects. In the scope of technical enhancements, advancing automation in the security patch management process, for example, automated detection of faulty patches \cite{dunagan2004towards, crameri2007staged, maurer2012tachyon} and mechanisms for reducing system downtime in reboots \cite{potter2005reducing, dumitracs2009upgrades, araujo2020improving}, have been widely studied. However, the literature presents little empirical evidence of the socio-technical aspects relating to security patch management. Existing socio-technical studies have primarily focused on the workflows of system administrators but did not focus on other roles (e.g., change manager) and external stakeholders (e.g., customer) involved in the patch management process. Crameri et al. \cite{crameri2007staged} were among the first to investigate system administrators' patch management practices. From a survey conducted with 50 system administrators, they reported that 70\% of administrators avoided deploying patches due to issues caused by a lack of integration between patch testing, deployment and post-deployment issue reporting. Dietrich et al. \cite{dietrich2018investigating} explored the system administrators’ perspective on factors leading to security misconfigurations. Their findings confirmed that the situation has not changed even after a decade, reporting that delaying and avoiding security patches are among the most frequently reported security misconfigurations. However, these studies did not explain the reasons for such delays or missed patches. 

Extending the study by Crameri et al., two recent studies \cite{li2019keepers, tiefenau2020security} have examined a larger sample of system administrators through a combination of surveys and interviews to perform a comprehensive investigation of the patch management process. Both studies explored system administrators' practices, behaviour, and experiences in the patch management process. According to them, administrators rely on various sources such as security advisories, direct vendor notifications, patch management tools, mailing lists and online forums to retrieve meaningful patch information. Further investigating system administrators' patch information retrieval-related needs and practices, Jenkins et al. \cite{jenkins2020anyone} studied how the mailing list of the website PatchManagement.org extends support in patch management activities. They argue that the mailing list acts as an online community of practice extending support not only in the patch information retrieval phase but throughout the process in various aspects such as guidance for patch prioritisation, workarounds for post-deployment issues and tool selection. 
  
Another set of studies \cite{li2019keepers, tiefenau2020security, dissanayake2021software, nappa2015attack, huang2012patch, potter2005reducing} has explored the challenges in the patch management process. For example, the impact of organisational policies and culture \cite{li2019keepers, tiefenau2020security, nicastro2003security}, collaboration and coordination challenges due to conflicts between stakeholders \cite{nappa2015attack, li2019keepers, huang2012patch, potter2005reducing}, lack of resources in terms of skills and expertise required for handling complex patching tasks \cite{post2003computer, tiefenau2020security, jenkins2020anyone}, and the increasing rate of patch release \cite{post2003computer, tiefenau2020security, potter2005reducing} are some of the most common challenges faced by practitioners. In addition, challenges relating to the lack of dedicated patch testing environments \cite{li2019keepers, maurer2012tachyon, tucek2009efficient}, post-deployment patch verification \cite{chen2014identifying, li2019keepers}, and system downtime during patch deployment \cite{li2019keepers, tiefenau2020security, potter2005reducing, dumitracs2009upgrades} have been widely discussed. Despite widespread attempts to the adoption of automation in different phases of the process, it is revealed that the need for human interaction still presents an inevitable challenge \cite{post2003computer, dunagan2004towards, li2019keepers, tiefenau2020security, dissanayake2021software}. 

To address some of these pressing socio-technical challenges, several studies have proposed tools, frameworks and practices. For example, a set of studies \cite{cavusoglu2006economics, cavusoglu2008security, dey2015optimal} has proposed synchronising an organisation’s patch cycle with the vendor’s patch release cycle to optimise the process, minimise stakeholder conflicts and reduce costs. Concerning the coordination challenges, Dissanayake et. al \cite{dissanayake2021grounded} have proposed a grounded theory of the role of coordination in security patch management explaining the causes that create the need for coordinating in the security patch management process, constraints for effective coordination, breakdowns resulting from ineffective handling of the coordination causes and constraints, and mechanisms for managing the causes while mediating the constraints. Although several approaches have been proposed to improve the patch management process to reduce delays, the reasons why such delays occur and how to mitigate them remain unexplored. Furthermore, given patch management is largely an industry-centric topic, relatively little has been done to understand the state of practice. For example, why do practitioners continue to delay applying the security patches leading to compromises that would have been easily prevented like the Equifax case \cite{Equifaxbug}?

In contrast, delays have been widely studied in related fields like software development. In the majority of these studies \cite{919083, 10.1145/358916.359003, 1205177, nguyen2008global}, the focus has been on delays in global software development (GSD). For example, they have explored the effect of distance on delays in a multi-site software development organisation and mechanisms to reduce delays. % The results have shown that cross-site work takes longer than comparable single-site work and requires more effort to coordinate the collaborative work. 
Closely related to our study but focused on software development projects is the empirical analysis conducted over a decade ago by Genuchten \cite{van1991software}. By analysing the planning data of six projects in one software development department of an organisation, he provided a classification of reasons for delays in software development activities. The findings report that capacity-related reasons cause the majority of the delays in the studied context. Further, the study highlights the importance of understanding the causes of delays for software developers to take necessary actions for improvement. Similarly, the recent expediting attacks targeting unpatched software security flaws exhibit a pressing need towards understanding the practical causes of delays in patch management and suitable strategies to take appropriate actions, which is accomplished by this study.
% Responding to this gap, we present \todo{a descriptive conceptual framework} that explains why delays occur when applying the security patches and the strategies that can be applied to reduce delays, grounded in practice. Further, we provide insights into the distribution of delays over the patching process and the distribution of the reasons for delays.

\section{Research Method}
\label{section:researchmethod}

\begin{figure*}[h]
  \centering
  \includegraphics[width=0.95\linewidth]{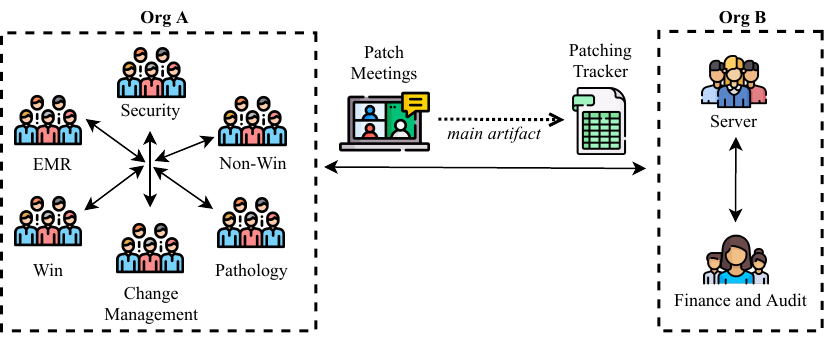}
  \caption{The organisational setup present in the studied context.}
  \Description{An overview of the organisational setup present in the studied context.}
  \label{fig:orgsetup}
\end{figure*}

To understand why and how delays occur in practice, we conducted a longitudinal study with two organisations (Org A and B) involving 21 participants from 8 teams in Australia. % This study was conducted under the University of X Human Research Ethics Committee Application Id Y. Abiding by the guidelines, the details of the companies, participants, third-party vendors, and customers have been kept confidential. 
In selecting the case organisations, we used a combination of purposive \cite{schwandt1997sage} and convenience \cite{marshall1996sampling} sampling to ensure that our data are representative of the substantive area through which the findings emerge. % This is because, in Grounded Theory, the findings are not generalized to a population, as the findings are relevant to the contexts studied \cite{glaser1967discovery, glaser1978theoretical}.
The demographics of the studied teams are illustrated in Table~\ref{tab:observationdetails}. Org A is an Australian state government health services agency that outsourced its \textbf{OS security patching} to Org B, an American multinational corporation. Org B was responsible for patching Org A's 1500 servers representing the entire health sector in the state government. In addition, the non-security patches were handled by different teams in Org A, which is not included in our analysis. Org A consisted of several teams each managing different modules, for example, the teams T1-3 represented the main modules in Org A while the teams T5-6 were central to all other in-house teams overseeing their respective modules. The security patch management process was coordinated through bi-weekly patch meetings between the two organisations, attended by key stakeholders representing each team detailed in Table~\ref{tab:observationdetails}. Abiding by the human ethics guidelines, the details of the companies, teams and participants have been kept confidential. Figure \ref{fig:orgsetup} shows the organisational setup in the studied cases. 

\begin{table*}[h]
  \caption{Demographics of participants}
  \label{tab:observationdetails}
  \centering 
  \small
  \begin{tabular} {p{.1\textwidth}  p{.035\textwidth} p{.3\textwidth} p{.07\textwidth} p{.37\textwidth}}
    \toprule
    Organisation & Team & Team's Domain & Team Size \anote{$^*$} & Roles\\
    \midrule
    Org A & T1 & Electronic Medical Records (EMR) & 5 & Application Owner, System Administrator, Server Engineer, Server Manager \\
     & T2 & Digital Health Windows (Win) & 3 & Server Engineer, System Administrator, Application Services Manager  \\
     & T3 & Digital Health Non-Windows (Non-Win) & 2 & Unix Specialist, Server Engineer, System Administrator \\
     & T4 & Clinical and Pathology \newline{}Services & 1 & Pathology Server Engineer \\
     & T5 & Security & 1 & Security Advisor \\
     & T6 & Change Management & 1 & Change Manager \\
    \hline
    \addlinespace
     Org B & T1 & Server (Technical) & 7 & Server Engineer, Senior Server Engineer, Unix Engineer, Server Manager, Client Delivery Manager \\
     & T2 & Finance and Audit \newline{}(Non-technical) & 1 & Accounts Manager\\
    \bottomrule
  \end{tabular}
\smallskip
\parbox[t]{\textwidth}{ $^*$~The team size refers to the number of team participants in the patch meeting.}
\end{table*}
% }

\subsection{Data Collection}

We collected longitudinal data from patch meeting minutes maintained by the case organisations as the main artefact from patch meetings. In addition, we observed the patch meetings to supplement our understanding of the process, activities, and strategies documented in the meeting minutes and verify the emerged findings from the artefact analysis. 

\begin{figure*}[h]
  \centering
  \includegraphics[width=1.0\textwidth]{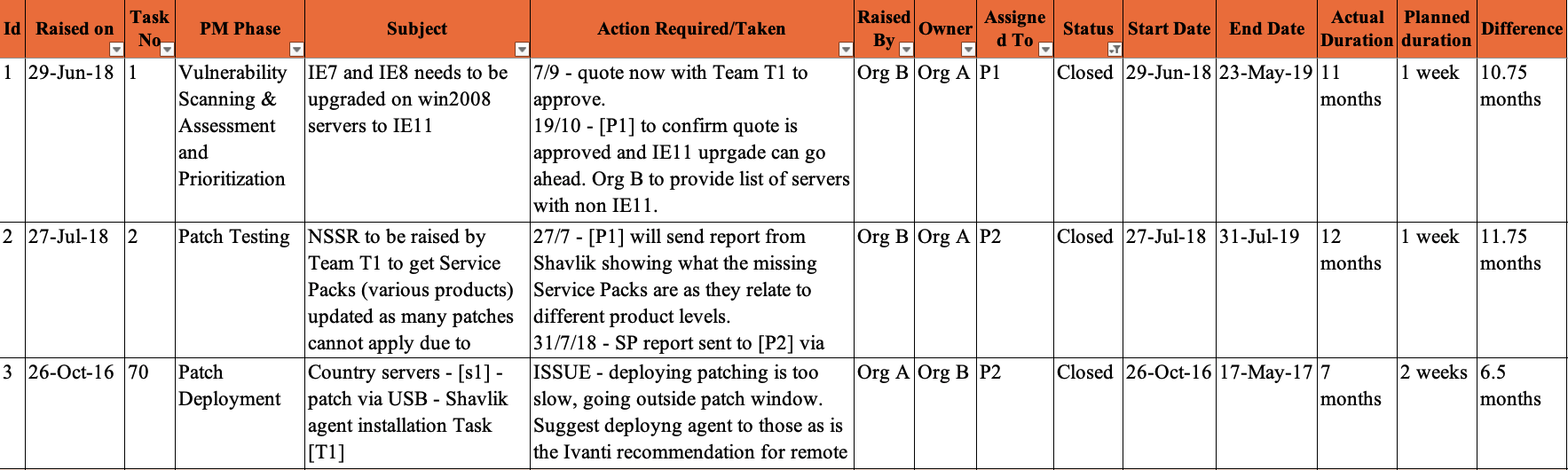}
  \caption{A screenshot of an extract from the Patching Tracker - 19.05.2021.} 
  \Description{An extract from Team T1's Patching Tracker on 19.05.2021.}
  \label{fig:patchingtracker}
\end{figure*}

% \textbf{artefact analysis.} 
As the main source of data collection, we gathered patch meeting minutes referred to as the \textit{``patching tracker"} by the studied teams. The patching tracker, a detailed Excel spreadsheet, was used as a tracking tool between the collaborative parties to document the status of the tasks, similar to a centralised version control and issue tracking system. The three main teams of Org A (i.e., T1, T2, and T3) each maintained separate patching trackers to document their patch management tasks (i.e., activities) with details of the task number, subject, raised date, action required or taken, raised by, owner, assigned to and the status (including \textit{Closed}, \textit{In-progress}, \textit{New}, \textit{On-hold} and \textit{Monitor}), as shown in Figure \ref{fig:patchingtracker}. Each tracker was updated regularly with the date and action or decision taken when the task was discussed in detail at patch meetings. 

% \textbf{Observations of patch meetings.} 
Additionally, to understand what occurs in practice and to obtain a better understanding of the documented tasks in the patching tracker, we observed 66 patch meetings from March 2020 - May 2021. The meetings provided a collaborative platform for the participants to discuss and refine the patching process, plan monthly patch schedules, assess the progress, resolve problems, and make decisions about patch exemptions. The fortnightly meetings were held online through Microsoft Teams due to COVID-19 and lasted approximately an hour and a half. Additionally, the observations increased analytical validity and ensured triangulation in our findings \cite{yin2009case}. 
% The first author participated in all 66 meetings and shared the artefacts among all authors following the meetings. We continued with the data collection until all authors mutually agreed that we have reached \textit{theoretical saturation} \cite{strauss1998basics}, i.e., the last few artefacts from patch meetings 57 to 66 provided no new insights, concepts, or categories but more examples and evidence.

\subsection{Data Analysis}

First, we qualitatively analysed the data employing the Grounded Theory's (GT) \cite{strauss1998basics, glaser1967discovery} data analysis procedures, particularly Strauss and Corbin’s version of GT procedures (\textit{Straussian GT}) \cite{strauss1998basics}, as they offer well-structured and rigorous data analysis techniques well-suited to answer complex and practice-based \textit{why} and \textit{how} type questions \cite{strauss1998basics}. Second, to identify how the delays and the causes are distributed, we quantitatively analysed the data using frequency analysis, a widely used technique for producing descriptive statistics derived from the data. 
% We followed Strauss and Corbin’s version of GT (\textit{Straussian GT}) \cite{strauss1998basics} because it provides a structured approach to data analysis with the research question(s) formulated upfront driving the research direction \cite{coleman2007using, kelle2007emergence, strauss1998basics}. % GT was well-suited to our goal of investigating and understanding why delays happen in security patch management in industrial practice as GT enables the investigation of people and interactions in real-world settings \cite{glaser1978theoretical}. Additionally, since GT enables the systematic development of theory from data to particularly explain \textit{why} and \textit{how} aspects of the underlying real-world phenomenon \cite{glaser1967discovery, glaser1978theoretical, strauss1998basics}, GT is considered most appropriate to research areas that have not been largely explored before \cite{hoda2011impact}. As the existing research on the delays of security patch management is scarce, the choice of GT as our research method was well justified. 

We analysed the data at the task level as previous studies \cite{van1991software, brooks1975essays} have shown that most project delays are caused by delays in the smallest unit of work (i.e., task-level delays). A task in this study refers to a single row recorded in the patching tracker. Out of 268 tasks available in total, 232 tasks were closed. We only analysed closed tasks since we needed the end dates to calculate delays. Figure \ref{fig:patchingtracker} shows a screenshot of an extract from the patching tracker. % We added six new columns to each Excel sheet, as shown in Figure \ref{fig:patchingtracker} in pink, comprising the ID, phase of the patch management process to which the task belongs, start date, end date, task duration, and reason for excluding the task for consideration if applicable.
To define a delay according to the studied context, the first author held a discussion with Org A's Security Advisor about their policies to understand the defined time frames for any given task during the monthly patch cycle practised. Table~\ref{tab:definitionoftimeframes} presents a summary of the standard time frames as expected in the organisation. Correspondingly, we mapped the tasks to their relevant phase of the patch management process based on two existing studies \cite{li2019keepers, tiefenau2020security}. 

%Table 1
% {\footnotesize\sffamily
% \setlength\tabcolsep{10pt}
\begin{table*}[h]
  \caption{Definition of standard time frames in the studied organisation}
  \label{tab:definitionoftimeframes}
  \centering 
  \small
  \begin{tabular} {p{.035\textwidth} p{.35\textwidth} p{.1\textwidth}  p{.4\textwidth}}
    \toprule
    Phase ID & Patch management process phase & Standard time frame & Note \\
    \midrule
    P1 & Patch Information Retrieval & 2 days & Needs to be completed within two days of patch release \\
    P2 & Vulnerability Scanning, Assessment and Prioritisation & 1 week & Needs to be completed within the first week of patch release \\
    P3 & Patch Testing & 1 week & Needs to be completed within the second week of patch release \\
    P4 & Patch Deployment & 2 weeks & Needs to be completed within the fourth week of patch release \\
    P5 & Post-Deployment Patch Verification & 1 month & Any post-deployment issues must be resolved by the next patch cycle \\
    \bottomrule
  \end{tabular}
\end{table*}
% }

The preliminary analysis revealed 132 delayed tasks from a total of 232 closed tasks that we analysed (56.9\%). While there were 57 tasks (24.6\%) not delayed, the remaining tasks were excluded for several reasons such as duplicate tasks, lack of information (e.g., no end date), and not being related to patch management specifically.

\begin{figure*}[h] 
  \centering
  \includegraphics[scale=0.7]{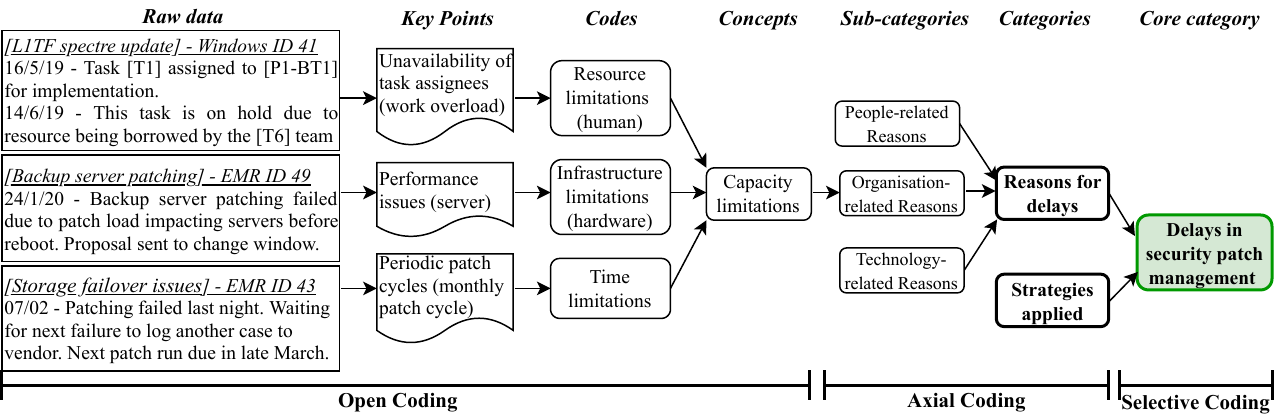}
  \caption{Emergence of the category \textit{Reasons for delays} from the underlying concept of \textit{Capacity limitations} and codes.}
  \Description{Application of Straussian grounded theory data analysis procedure: open coding through axial coding to selective coding.}
  \label{fig:dataanalysis}
\end{figure*}

To understand the causes of delays and remediation mechanisms, we analysed in-depth the delays identified through preliminary analysis following \textbf{open}, \textbf{axial}, and \textbf{selective} coding procedures \cite{strauss1998basics}, as shown in Figure \ref{fig:dataanalysis}. The first author performed the data analysis while the second and third authors cross-checked all the codes throughout the process to increase the reliability of the findings and reduce bias \cite{strauss2007basics}. Any disagreements in the coding were resolved through weekly discussions among all authors involving multiple rounds of revisions. The patching tracker, codes, and memos were stored in NVivo, the data analysis tool, and shared with all authors.

We started with \textbf{open coding}, whereby we analysed all columns row by row in the spreadsheet to identify \textit{key points} summarising the content. It was further summarised into \textit{codes} containing short phrases. Constant comparison of emerged codes between each team's patching tracker, different teams' patching trackers of a single meeting, and different meetings resulted in \textit{concepts} \cite{strauss1998basics}, a higher level of abstraction of the \textit{codes}. Similarly, we grouped \textit{sub-categories}, and continuously comparing sub-categories gave rise to \textit{categories}, the next level of abstraction. Next, we performed \textbf{axial coding} in which we linked categories to their subcategories based on the relationships between categories relating to their \textit{properties} (i.e., ``characteristics of a category") and \textit{dimensions} (i.e., ``variations within properties") \cite{strauss1998basics}. During the analysis, we created \textit{memos} for explaining the codes and their relationships which helped us during this process.
% Using \textit{memos} (or \textit{memoing}), i.e., notes written by analysts explaining the codes and relationships during the data analysis, helped us during this process. 
During the final phase of the analysis, we applied \textbf{selective coding} by which we identified the \textit{central} or \textit{core} category which represents the most recurrent and central problem in the studied phenomenon, or simply which explains \textit{``what this research is all about”} \cite{strauss1998basics}, in this case, \textbf{delays in security patch management}. 

For confidentiality reasons, we do not share the raw data. However, we made our codebook containing the codes, descriptions and examples of raw data publicly available \footnote{\url{https://doi.org/10.5281/zenodo.5635608}}. 

\subsection{Member checking}

We conducted a member checking \cite{merriam1998qualitative} session to ensure the credibility, accuracy, validity, and transferability of our study findings. Member checking, a technique of \textit{``taking ideas back to research participants for the confirmation"} \cite{charmaz2006constructing}, provides an opportunity to validate the findings with participants and resonance with their experiences \cite{birt2016member}. We presented the study findings at a session held at Org A. Three authors attended the session in person while nine patch meeting participants (six from Org A and three from Org B) and an executive director of Org A were present physically. In addition, seven patch meeting participants (four from Org A and three from Org B) attended the session virtually. The first author presented the findings for 20 minutes followed by a detailed feedback discussion lasting for 40 minutes. For the member checking, we revisited findings for each RQ and asked questions including if they agree with the findings, which reasons for delays they have encountered the most in their experience, any other reasons or strategies they use that are not captured in the findings, and if they can relate the findings with their experiences. The session was audio-recorded with permission and transcribed for analysis by the first author. The feedback and comments from member checking are presented in Section 4.5.

\section{Findings}
\label{section:findings}

\begin{figure*}[ht]
  \centering
  \includegraphics[width=1\textwidth]{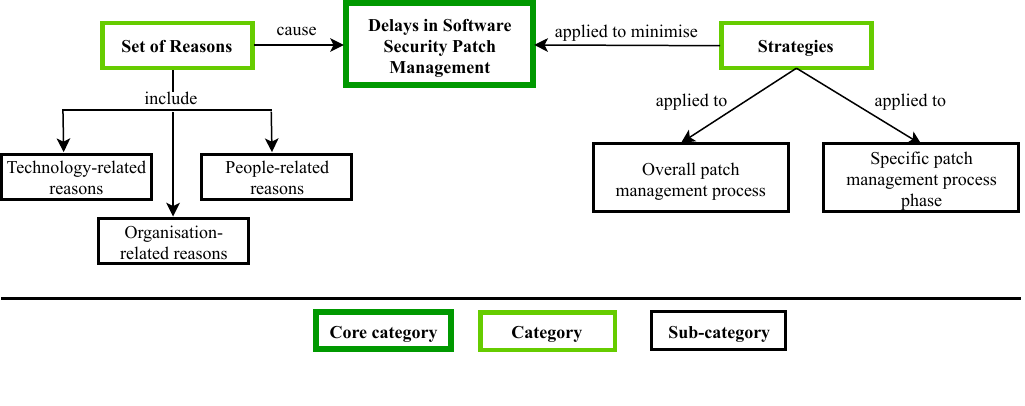}
  \caption{High-level overview of the findings from the qualitative data analysis.}
  \Description{High-level overview of the reasons and strategies for delays in software security patch management emerged from a Grounded Theory analysis.}
  \label{fig:theory}
\end{figure*}

In this section, we present the findings of our study. Figure \ref{fig:theory} presents an overarching representation of the findings from the qualitative analysis. We provide examples from the patching tracker chosen based on their representativeness, as supporting evidence and to increase the verifiability of our findings \cite{urquhart2012grounded}. In the examples, we include the subject of the task (see Figure \ref{fig:patchingtracker}) and evidence relating to the delay using unique identifiers for ease of reference, for example, \textit{``P[n]-AT1"} refers to a participant from Org A’s EMR team, and \textit{``Win, Task ID 2"} refers to the 2nd task discussed in the Digital Health Windows meeting. 
% We also provide the number of references (in parentheses) to represent their frequency of occurrence.
% Further, we present the results from a quantitative analysis providing insights on the \textbf{distribution of the reasons for delays (RQ3)} and \textbf{distribution of delays over the patch management process (RQ4)} that will be useful in identifying \textbf{when} and \textbf{what} in the patch management process needs actions for improvement to reduce delays. 

\subsection{Why, how, and where do delays occur in security patch management?}

\begin{figure*}[ht!]
  \centering
  \includegraphics[width=1\textwidth]{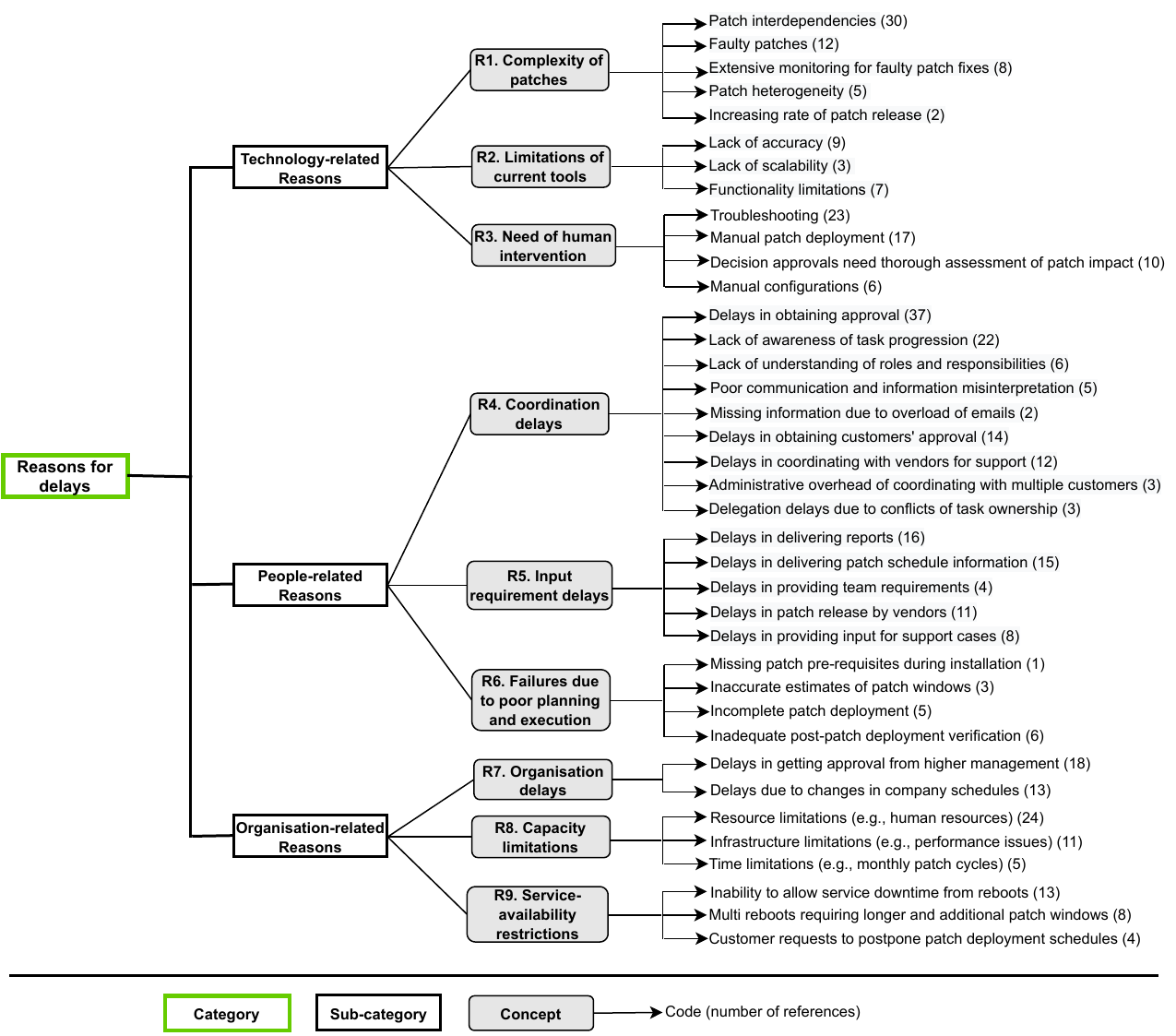}
  \caption{Detailed overview of the causes of delays in software security patch management.}
  \Description{Reasons for delays in software security patch management.}
  \label{fig:detailreasons}
\end{figure*}

\begin{figure}
    \centering
    \begin{minipage}{0.49\textwidth}
        \centering
        \includegraphics[width=1\textwidth]{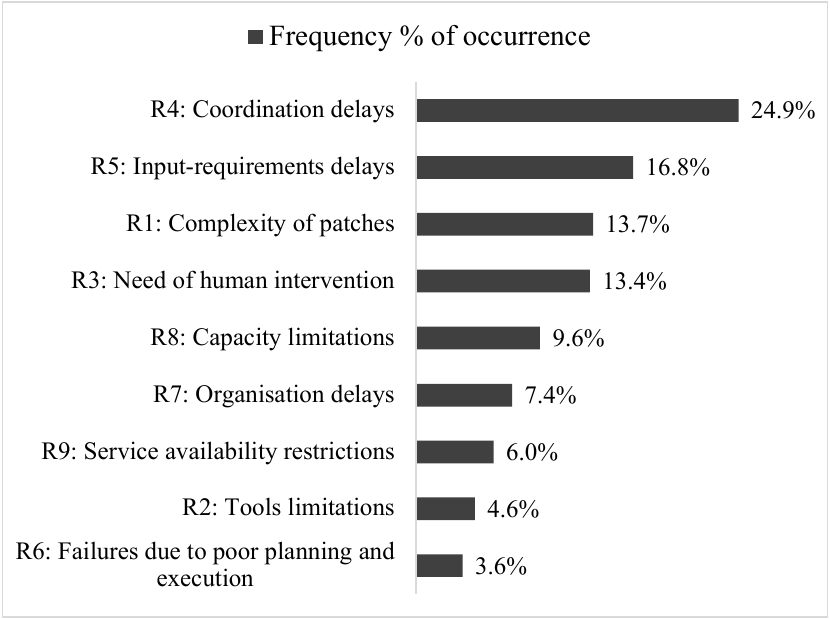}
        \caption{Frequency distribution of the reasons for delays in security patch management from a total of 417 occurrences of delayed reasons.}
        \Description{Distribution of the reasons for delays in software security patch management based on the frequency analysis.}
        \label{fig:delaysfrequency}
    \end{minipage}\hfill
    \begin{minipage}{0.49\textwidth}
        \centering
        \includegraphics[width=1\textwidth]{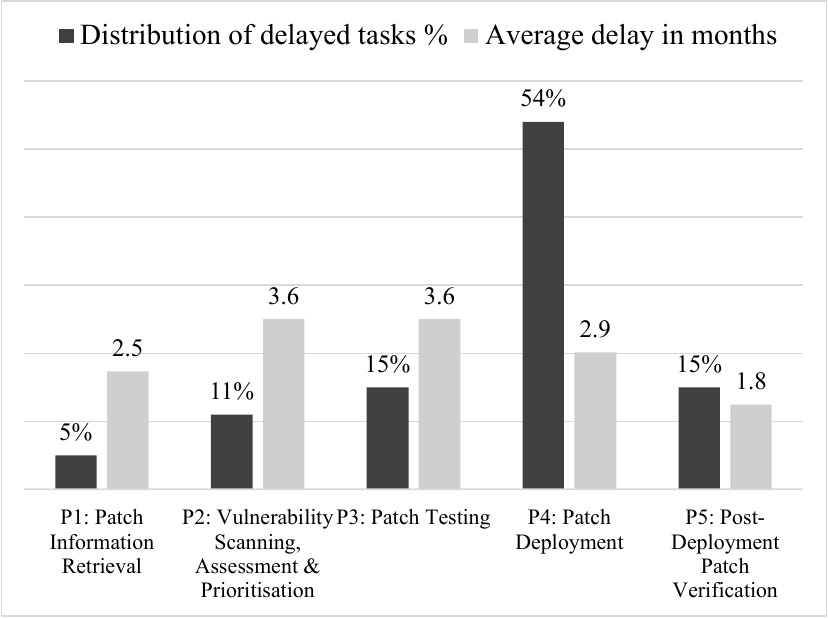}
        \caption{Distribution of delays over the security patch management process and average delay duration in months in each phase. Total number of delayed tasks = 132.}
        \Description{An overview of how the delays are distributed over the patch management process and the average duration of the delays in each process phase.}
        \label{fig:delaysdistribution}
    \end{minipage}
\end{figure}

We identified a set of reasons that cause delays in security patch management, presented as a taxonomy in Figure \ref{fig:detailreasons}. In summary, we found nine reasons, grouped into three main categories: technology-related reasons, people-related reasons and organisation-related reasons.
% \subsection{RQ1.2. How often do these causes of delays occur?}
Next, we quantitatively analysed the identified set of reasons to understand the \textit{most prominent reasons} that need practitioners' and researchers' attention. To achieve this, we conducted a frequency analysis on the reasons for delays. An important observation was that in the majority of the delays, we found multiple reasons attributing to one delayed task. For example, a delay in applying a critical security patch was identified due to a combination of reasons such as delayed input by the vendor (R5), delays in coordination with the vendor (R4), and lack of expertise (R8). In total, we found 417 occurrences of the identified nine reasons ascribed to the 132 delayed tasks analysed. Figure \ref{fig:delaysfrequency} presents the frequency distribution of the reasons for delays. Accordingly, the most prominent causes for delays relate to people-related reasons, for example, delays in coordinating the patch management activities (24.9\%) and providing input requirements (16.8\%). % They are closely followed by technology-related reasons, i.e., the complexity of patches (13.7\%) and the need for human intervention (13.4\%).

% \subsection{RQ1.3. Where do delays occur in the patch management process?}
In determining \textit{where the aforementioned delays occur in the patch management process}, our quantitative data analysis revealed that the delays are distributed throughout the process with a majority of the delays, i.e., 54\% occurring during the patch deployment (P4) phase as shown in Figure \ref{fig:delaysdistribution}. We identify that it can be attributed to the inherent socio-technical complexities involved in the patch deployment tasks and decisions. The second-highest number of delays happen during patch testing (P3) and post-deployment patch verification (P5) phases where each account for 15\% of the delays. Possible explanations of these numbers can be recognised by the evident challenges in the respective stages, for example, managing the delays occurring due to the poor quality of patches, which may result in unanticipated post-patching failures leading to disastrous consequences and inconvenience to users, e.g., unavailability of service. Additionally, we have reported the average delay duration in months in each process phase. As shown in Figure \ref{fig:delaysdistribution}, the longest average delay is reported as 3.6 months belonging to patch testing and vulnerability scanning, assessment and prioritisation tasks. In the following, we describe the nine reasons for delays mentioned in \textit{italic} under their corresponding main categories.

\subsubsection{\textbf{Technology-related reasons.}} The technology-related reasons denote the compound characteristics intrinsic to software security patches, limitations of the tools used in patch management, and technological limitations resulting in the need of human intervention in the process. \\

Concerning the \textit{complexity of patches}, the patch interdependencies consisting of software, hardware, and firmware presented a major reason for delays in patch testing and deployment tasks. We identify that such complexities emerge from the existing dependencies in the source code, for example, function-level or library-level dependencies \cite{dissanayake2021grounded}. Patching large and complex software systems involves a diverse set of operating systems, tools, and software applications with multiple versions. It introduces additional challenges to match the compatibility of several versions which often leads to delays during patch testing.
% \begin{quoting}
% \textit{\emph{[Subject - Software version issues on the system [EPAS 17.3] to resolve]} \\
% ``24/1/19 - 47 Office patches on servers [s1-6] (multi-languages and x32 and x64 bit versions) and 7 Visual Studio-related security patches having errors while testing. \\
% 6/2/19 - A ticket raised to investigate Visual Studio patches and manually apply if required."} - EMR, Issue ID 9
% \end{quoting}
Moreover, patch interdependencies with the legacy software were a recurrent cause of the delays in the studied context. This reason exacerbated problems with delays since the solutions, for example, upgrading or decommissioning the legacy system, or continuing to receive extended support (i.e., obtain patches) from the vendors presented even further challenges. This is because, besides the large costs involved in these workaround solutions, the teams were faced with high risks as most of the legacy systems operated on critical medical services. In addition, due to the complex and business-critical nature of legacy systems, resolving legacy software dependencies often resulted in significant delays leading up to several months in some cases.
% \begin{quoting}
% \textit{\emph{[Subject - .NET Core patching]} \\
% ``4/9/20 - Patch deployment failed because a version upgrade was required to apply the patch. Need to reschedule the failed deployment."} - EMR, Issue ID 9
% \end{quoting}

On the other hand, the unknown errors during patch testing, deployment, and post-deployment arising from faulty patches led to delays. In such instances, the practitioners spent a significant amount of time troubleshooting the error not knowing that it is caused by a faulty patch. Following the identification of the root cause as a faulty patch, the practitioners often pursued the vendor's support which further delayed the completion of the task. 
% \begin{quoting}
% \textit{\emph{[Subject - [Server s1] non-functional after patching]} \\
% ``30/10/18 - Org A raised a support case to the [vendor v1] as the server was left in an unusable state with no communication post-patching.\\
% 4/4/19 - No further update regarding the support case."} - Win, Issue ID 6
% \end{quoting}
We also observed that some security patches required extensive monitoring to verify the fixes for post-deployment errors. For example, the task was kept under monitoring for several weeks until the results confirmed the applied fix poses no unanticipated adverse effects to the managed systems. Furthermore, the increasing rate of patch release coupled with the patch heterogeneity adds to the complexity of patches creating delays in patching. This is because as the number and diversity of patches increase, the number and complexity of the patch interdependencies that need to be managed also increase. Consequently, it leaves myriad attack vectors vulnerable to cyberattacks increasing the risk of exploits.
\begin{quoting}
\textit{\emph{[Subject - Patch deployment error at the [server s1]} \\
``13/12/19 - Workaround applied and timings were all good. Keep open till January run for confirmation."} - EMR, Issue ID 35 \\
\end{quoting}

The analysis unveiled that some delays can be attributed to the \textit{limitations of tools}. In particular, the lack of accuracy in the output of current tools (e.g., missing some vulnerabilities during scanning, omitting patches during patch deployment) resulted in inaccurate vulnerability prioritisation and incomplete patch deployment respectively. Subsequently, the practitioners had to re-execute the tasks resulting in delays in the task completion.  
% \begin{quoting}
% \textit{\emph{[Subject - Issues with IE patches]} \\
% ``7/8/19 - [...] The Windows update is showing zero missing patches but the [patch deployment tool] is showing 15 missing patches related to IE. [P1-AT2] to raise a support case for Microsoft account manager attention."} - EMR, Issue ID 35
% \end{quoting}
Another limitation is associated with the lack of scalability to handle diverse types of patches and their features. % For example, the new security feature introduced by Microsoft to protect the authenticity of the patch requires decryption of the SHA-1 and SHA-2 hash algorithms before the patch download. 
In such cases, patches introduce complications to tool functionalities such as disabling some tool functions. Furthermore, we identified functionality limitations of existing tools like the inability to detect patch compatibility arising from the patch dependencies and the lack of capability to detect multi-reboot requirements that delayed the tasks.   
\begin{quoting}
\textit{\emph{[Subject - Additional reboot required for .NET patching]} \\
``7/2/20 - An investigation is needed around the number of required reboots for EMR patching and window requirements as a result if more reboots are required. A new process needs to be fleshed out when patching is postponed to accommodate the identification of the number of reboots required."} - EMR, Task ID 35 \\
\end{quoting}

Another prominent cause of delays is ascribed to the \textit{need of human expertise} throughout the patch management process. The need for human intervention emerges because of the inability to achieve complete automation in the process owing to technological limitations. Troubleshooting the issues, mostly related to the unknown errors during and post-deployment, and faulty patches consumed a lot of the practitioner’s time and effort delaying patch testing and deployment tasks. 
% \begin{quoting}
% \textit{\emph{[Subject - Issues with patch installation at [server s1]]} \\
% ``17/4/19 - Installed patches reverting and some patches not getting installed, need further investigation. \\
% 3/5/19 - [P1-BT1] ran troubleshooting, no success. Need to raise a support case with Microsoft for advice."} - Win, Issue ID 20
% \end{quoting}
Similarly, manual configurations, for example, selecting the suitable Group Policy Object (GPO) configurations based on the needs and making decisions about the patch process, e.g., changes to the patch cycle and patch window, needed to be thoroughly assessed for the impact on multiple aspects to avoid breakdowns. Moreover, we noticed that the practitioners undertook manual patch deployment during complex, erroneous, or business-critical patch installations, for example, legacy systems patching. Manual intervention was also required for re-executing failed patch deployments and re-planning patch schedules due to requirement changes. 
\begin{quoting}
\textit{\emph{[Subject - [Hospital h1] patching stage 3 on 27th November]} \\
``18/10/19 - Patching needs to be moved to OOB due to the change freeze from 15th November to 3rd December. \\
31/10/19 - [B-T1] team putting in significant amounts of work, like 15-20 hours per month, to redo the schedules on custom dates each time the deployments move off standard windows."} - EMR, Task ID 30
\end{quoting}

\subsubsection{\textbf{People-related reasons}.} These refer to a group of reasons relating to the coordination of patch management, delivery of input requirements, and planning and execution of patch management tasks. \\

Delays occurring due to \textit{lack of coordination} presented the most recurrent reason for delays. It refers to the delays in getting things done in the patch management process. It is challenging because completing a single patching task (e.g., applying security patch X to server Y in Customer Z) involves multiple interdependent activities and several stakeholders. We found coordination delays stemming from both internal and external stakeholders.
    
\hspace{4 mm} Internal stakeholder coordination delays, in the studied context, relate to the delays from lack of coordination of dependencies deriving from the interactions between stakeholders of Org A and Org B. As several interdependent teams between the two organisations collaboratively worked towards an end goal of timely application of security patches to ensure systems' security, a delay of one party resulted in delays in task completion. Similarly, a lack of awareness of task progression between teams also created delays in inter-team task progression.
% \begin{quoting}
% \textit{\emph{[Subject - Need Service Packs updated to resolve software version incompatibility.]} \\
% ``31/8/18 - Not sure if [A-T1] team has raised Non-Standard Service Request (NSSR) ticket to upgrade 32-bit clients that are lower than IE8. \\
% 14/9/18 - Well, [P1-AT1] must know but he's not at the meeting, I'm not sure about it."} - Win, Task ID 2
% \end{quoting}
As such, the multiteam system \cite{mathieu2001multi} in the studied context resulted in delays in decision approvals as they had to go through multiple teams (or levels). 
% \begin{quoting}
% \textit{\emph{[Subject - IE version upgrade approval required]} \\
% ``7/9/18 - quote now with [A-T2] team to approve. \\
% 3/5/19 - [P1-AT4] to follow up with [P2-AT2] on the status of this. \\
% 10/5/19 - [P1] following up, still waiting for approval."} - Win, Task ID 1
% \end{quoting}
In addition, a lack of understanding of shared roles and responsibilities led to delays in coordinating tasks between teams because the task assignee did not know whom to contact in the event of errors or who was handling the interdependent task.
% \begin{quoting}
% \textit{\emph{[Subject - Investigation of post-deployment issue]} \\
% ``18/10/19 - [P1-BT1] says Org B should not be accountable for apps that don't start correctly. \\
% 31/10/19 - Decided that Org A or third-party vendor X should be fixing this error, Org B should not be accountable for application problems."} - EMR, Issue ID 26
% \end{quoting}
We noticed that coordination delays also occurred due to missing information owing to an overload of emails. Email being the primary source of communication between the internal teams, there were cases where some emails had been missed resulting in delays in passing information on time. Moreover, poor communication and information misinterpretation contributed to delays in information passing.
% For example, delays attributed to misinterpretation of communication and the frequent need to constantly follow up to get responses.
% \hspace{1 mm} \textit{``\emph{[subject - NSSR to be raised for investigation of effort and planning for pre-downloading for auto-patching.]} \\
% 15/11/19 - [P1's] advice [for this task] was misunderstood and confused with multi-reboot.  Pre-download has now been scheduled for the entire fleet. Item can close."} 
% \begin{quoting}
% \textit{\emph{[Subject - Cluster-based patching (Type 4a vs Type 4 .NET patches)]} \\
% ``12/7/19 - miscommunication regarding Types 4a and 4. \\
% 9/8/19 - Type 4a is now named Type 4-EMR. [P1-AT1] to send out email to advise wider audience."} - EMR, Issue ID 18 \\
% \end{quoting}
    
\hspace{4 mm} Concerning external stakeholder coordination delays, we found delays attributed to the coordination with customers (e.g., hospitals), end-users (e.g., hospital patients and staff), and vendors (e.g., Microsoft). A dominant reason was the delays in obtaining customers' approval for patch deployment. Since patch deployment usually resulted in system downtime arising from the reboots, obtaining approval for patch deployment schedules was important. % However, we observed significant delays in patching due to the delays in getting customers' approval.  
\begin{quoting}
\textit{\emph{[Subject - Request to change patch window of [s1] server]} \\
``10/5/19 - Currently set to 0000-0300, but the full backup of the server happening during this window causes slowness and issues with patching. Suggest changing the window to 0600-0900. [P1-AT2] checking on the status with business approval \\
14/6/19 - [P1-AT2] to follow up as no response from the business."} - Win, Task ID 19
\end{quoting}
The other reason was the delays in coordinating with vendors for support. Coordinating the vendor dependencies is integral to security patch management as the practitioners rely on vendors' support for errors encountered during patching and to obtain information about patch releases.
% \begin{quoting}
% \textit{\emph{[Subject - Patch deployment issue - file validation errors in [s1] server]} \\
% ``25/2/19 - [Vendor v1] case [n] - number of fixes suggested, completed last tests on server [s1], still no fix found. \\
% 3/5/19 - awaiting full response from [Vendor v1], investigations still underway."} - Win, Issue ID 17
% \end{quoting}
Additional delays included the administrative overhead of coordinating with multiple customers for pre and post-patching verification and delegation delays due to conflicts of task ownership with other third-party vendors owing to lack of accountability. \\
% \begin{quoting}
% \textit{\emph{[Subject - Verify client email contact for the task [T1]]} \\
% ``27/2/19 - Provided email contact gives an error message stating that the email is restricted. So [Org B] has altered the special instructions to send emails to [P1] until sorted. \\
% 1/3/19 - Org A to contact the client and verify the client's email address."} - Win, Issue ID 14 \\
% \end{quoting}
    
Another instrumental people-related reason was the \textit{delays in providing input requirements}. This is because the patch management process represents a sequence of phases with tightly coupled activities whereby an output of one phase is the input to the next phase. Similar to coordination delays, we identified that the input requirements delays emerge internally and externally. 

Internal input requirements delays occurred when requested information was not provided by the internal teams on time. This included delays in delivering the reports such as the vulnerability scan reports which led to delays in vulnerability assessment and prioritisation. Similarly, delays in delivering patch schedules-related information led to delays in planning and subsequently deploying patches. 
% Since Org B needed approval for all patch schedule plans from Org A prior to patch deployment, Org A's delays in providing information about patch cycle changes, patch windows, and organisational schedule changes such as change freeze deferred planning and execution of patch deployment by Org B. 
% \begin{quoting}
% \textit{\emph{[Subject - Scheduling Out-Of-Band (OOB) patching for exempted servers]} \\
% ``14/6/19 - An email sent to [P1-AT2] asking for patch window information, pending response. \\
% 28/6/19 - A follow-up email was sent to [P1-AT2] as no information was received. \\
% 26/7/19 - The patch window is still pending"} - Win, Issue ID 22
% \end{quoting}
Other reasons included delays in supplying other information requisites such as server details and providing the team's requirements in the patch cycle. An important observation was that the teams did not maintain an online repository with the server details which created the need for waiting for information about up-to-date server details (i.e., with the latest patched versions). 
% \begin{quoting}
% \textit{\emph{[Subject - IE version upgrades resulting in additional patches]} \\
% ``29/5/20 - [P1-AT4] waiting for the server names from [P2-AT1] for a fresh scan. [P2-AT1] to forward the full list to [P1-AT4] next week."} - EMR, Issue ID 44
% \end{quoting}
% Two other reasons identified were the delays in providing team requirements for patching tasks. Given the mix of environments and the multiple teams involved, a delay in delivering the requirements and specs of hardware created delays in subsequent planning and coordination. 
% \begin{quoting}
% \textit{\emph{[Subject - SQL Security Patching]} \\
% ``1/5/19 - No update since 17/4. Seeking confirmation on whether this is still a requirement from us [BT1]."} - Non-Win, Issue ID 1 \\
% \end{quoting}

\hspace{4 mm} External input requirements delays are concerned with the requirements delivered late by vendors. For example, delays in the patch release, particularly the patches for fixing critical security vulnerabilities, can result in a significant increase in the risk of exposure to cyberattacks. Additionally, the delays in receiving vendor's support for patching errors and new patch release information caused delays in addressing the vulnerabilities.
\begin{quoting}
\textit{\emph{[Subject - New zero-day vulnerability warning]} \\
``12/6/20 - Monitor Microsoft patch release for critical vulnerability identified on [T1] servers. Font Type 1 expected as a zero-day soon, full report not available yet. \\
24/7/20 - No update from Microsoft."} - EMR, Task ID 43 \\
\end{quoting}

We noticed that some delays were caused by \textit{failures from poor planning and execution}. Security patch management in large and mission-critical domains like healthcare entails challenging tasks that need to be cautiously planned and executed to avoid system breakdowns. However, the complexity of patches, particularly, the unforeseen errors during deployment presented a major risk to deploying within the planned time frame. With regards to poor planning, inaccurate estimates of patch windows caused patch deployment to exceed the allocated patch windows resulting in inconvenience to customers and end-users. As a consequence, practitioners often halted patch deployment to avoid service disruptions resulting in patching delays. 
\begin{quoting}
\textit{\emph{[Subject - Execution exceeding the patch window]} \\
``31/5/17 - Only 72.9\% of scheduled patch deployments were completed as of 11.20 am.  Two further windows to be raised to ensure the appropriate length of time is scheduled due to unknown 2016 updates that were required to be implemented, first window is 1st June 8 am to 12 pm."} - Win, Task ID 4
\end{quoting}
Of poor execution, missing patch prerequisites such as registry changes, GPO configuration and installation of preparation packages halted execution due to errors during the deployment. Similarly, incomplete patch deployment (e.g., failing to reboot after deployment which resulted in the installed patch not taking effect), and inadequate post-deployment patch verification such as failing to monitor the status of patch deployment tasks caused the need to re-execute patch deployment. Insufficient post-deployment patch verification also resulted in operation disruptions due to unexpected errors. For instance, we observed a heated discussion during a patch meeting owing to an issue with the printers not working reported by the customers to Org A caused by a lack of post-deployment verification by Org B.
% \begin{quoting}
% \textit{\emph{[Subject - [Server s1] post-patching functionality issues]} \\
% ``8/10/19 - [Server s1] did not shut down properly during the patch window which caused the printers to be unavailable for [customer c1]. It needs to move to semi-auto patching for November to monitor."} - Win, Issue ID 32 \\
% \end{quoting}
% \end{enumerate}

% \noindent\fbox{%
%     \parbox{\textwidth}{%
%         \textbf{Summary of RQ1.1:} We identified a set of reasons that cause the delays in software security patch management. We provided a taxonomy of the reasons comprising two technical reasons and seven socio-technical reasons for delays.
%     }%
% }

\subsubsection{\textbf{Organisation-related reasons}.} This category covers reasons relating to the organisation approvals, schedules, capacity to undertake patch management tasks and policies on service availability. \\

We found some reasons denoting \textit{organisation delays} resulting from organisation policies and schedules. The need for compliance with organisation policies and the involvement of multiple parties (i.e., two organisations and several teams) has resulted in delays in obtaining approval from organisational management for monthly patch schedules and changes in the process. 
% \begin{quoting}
% \textit{\emph{[Subject - RITM for new bespoke solution]} \\
% ``8/10/19 - Approval received from AT1, has been sent across to [P1-AT2] to raise the purchase order. \\
% 29/11/19 - The requested item is still with Finance (team) for processing"} - EMR, Issue ID 23 \\
% \end{quoting}
We also noticed that delays occur due to changes in organisation schedules such as change freeze periods, testing schedules like regression testing plans, and holidays (e.g., year-end shutdown period) during which no patch deployments were allowed to be scheduled.
\begin{quoting}
\textit{\emph{[Subject - Patching for December 2019]} \\
``18/10/19 - OOB for November patching from 4th December instead of December patching. \\
31/10/19 - [AT1] patching for December month is off but November Microsoft patches will be applied in the first week of December instead to keep compliance up."} - EMR, Issue ID 29 \\
\end{quoting}

Further, we noticed a \textit{lack of capacity} concerning human resources, infrastructure and time leading to delays. With regards to resource constraints, insufficient human resources appeared to be a major factor in delays. For example, unavailability of task assignees due to high work overload and assignee being on leave held up the tasks in progress until the assignee was available. Another root cause was the lack of qualified personnel with sufficient experience to handle complex tasks such as legacy system upgrades, thus leading to an experienced practitioner getting overloaded with tasks that would end up queued for a long time. 
% \begin{quoting}
% \textit{\emph{[Subject - L1TF spectre update]} \\
% ``16/5/19 - Task 0117374 assigned to [P1-BT1] (a senior server engineer) for implementation. \\
% 14/6/19 - Task on hold until July due to resource (i.e., P1-BT1) being borrowed by [AT6] team (for another critical task)."} - Win, Issue ID 21
% \end{quoting}
Regarding infrastructure-related limitations, hardware and network limitations hindered task progression in ways such as performance delays. For example, the high patch load described in the \textit{complexity of patches (R1)} impacted the reboots following deployment and issues with the bandwidth required for patching due to a lack of capacity to handle the load. 
\begin{quoting}
\textit{\emph{[Subject - Backup server patching]} \\
``24/1/20 - Patching cannot go ahead when the active backup is running. The patch load can impact servers before reboot. Need a window change, proposal to be sent by [P1-BT1] to [P2-AT1]."} - EMR, Issue ID 39
\end{quoting}

Another reason stemmed from the periodic patch cycles as it presented the practitioners with a time-bound restraint to progress with the tasks. In particular, some tasks such as testing the workarounds for failed deployments had to be delayed for weeks given the time-driven (i.e., monthly) patch cycle in practice. \\  
% \begin{quoting}
% \textit{\emph{[Subject - [Server s1] storage failover issue]} \\
% ``7/2/19 - A support case was raised with [vendor v1]. Waiting for the next failure to occur to submit the logs, the next patch run is due in late March."} - EMR, Issue ID 33 \\
% \end{quoting}

Another crucial cause of delays stemmed from the \textit{service availability restrictions}. We noticed that patch deployment was often delayed due to organisations' inability to allow service downtime from reboots. Reboots were necessary for the patch to take effect after deployment and some patches required multiple reboots or multi reboots % (i.e., reboots before, during, and after patch deployment) 
depending on the level of complexity involved, for example, the number of patch interdependencies. As such, the multi reboots required longer and additional patch windows than the usually allocated 4-hour window. Consequently, in most cases, the patch schedules were delayed to be deployed in out-of-band (OOB) windows to reduce service disruptions from longer patch windows during business hours. 
\begin{quoting}
\textit{\emph{[Subject - [Servers s1 and s2] patching]} \\
``26/7/19 - OOB window is needed for the multi reboots to catch up. \\
9/8/19 - Waiting for the customer's confirmation of the new patch window, pending information from [P1-AT1]."} - EMR, Issue ID 20
\end{quoting}

However, getting customers' approval for a change of patch window presented an additional challenge to the practitioners as customers were always hesitant about the risk of system downtime. Correspondingly, further delays occurred due to customers' requests to postpone the schedules to allow service continuity.\\
% \begin{quoting}
% \textit{\emph{[Subject - Go-Live for [hospital h1] on 19th July]} \\
% ``12/7/19 - Customer [hospital h1] asking to postpone patching until 7th August."} - EMR, Issue ID 16 \\
% \end{quoting}
% \begin{tcolorbox}[colback=white, left=0.5pt, top=0.5pt, right=0.5pt, bottom=0.5pt]
%     \textbf{Summary about why, how, and where do delays occur in security patch management}. We identified nine causes for patching delays associated with technology, people and organisation-related reasons. Among these reasons, we found people-related reasons (e.g., coordination and input requirement delays) as the most prominent and recurrent reasons. While the delays are distributed throughout the patch management process, most of the delays, i.e., 54\% occurred during the patch deployment phase. We also found that on average vulnerability scanning, assessment and prioritisation, and patch testing tasks were delayed the most, i.e., 3.6 months.
% \end{tcolorbox}

\noindent\fbox{%
    \parbox{\textwidth}{%
        \textbf{Summary for RQ1:} We identified nine causes for patching delays associated with technology, people and organisation-related reasons. In a majority of the delays, we found multiple reasons attributing to one delayed task. Among these reasons, people-related reasons, for example, coordination delays and input requirement delays appeared as the most prominent and recurrent reasons. Concerning where the delays occur, we found that the delays are distributed throughout the security patch management process, however, most of the delays, i.e., 54\% occurred during one phase, i.e., patch deployment. Yet, regarding the duration of delays, we found that tasks related to vulnerability scanning, assessment and prioritisation and patch testing phases account for the longest delays.
        % The causes for patching delays are associated with technology, people and organisation-related reasons. In a majority of the delays, multiple reasons can be attributed to one delayed task. Among these reasons, people-related reasons, for example, coordination delays and input requirement delays appear as the most prominent and recurrent reasons. Concerning where the delays occur, the delays are distributed throughout the security patch management process, however, most of the delays, i.e., 54\% occur during one phase, i.e., patch deployment. Yet, regarding the duration of delays, tasks related to vulnerability scanning, assessment and prioritisation and patch testing account for the longest delays.
    }%
}
% \noindent\fbox{%
%     \parbox{\textwidth}{%
%         \textcolor{mediumblue}{\textbf{Key Findings}:\\
%         • Identified nine causes for patching delays associated with technology, people and organisation-related reasons. \\
%         • Among these, the most prominent reasons are people-related (e.g., coordination and input requirement delays). \\
%         • The delays are distributed throughout the security patch management process, however, most of them (54\%) occurred during the patch deployment phase. \\
%         • On average, vulnerability scanning, assessment and prioritisation, and patch testing tasks are delayed the longest.}
%     }%
% }
 
\subsection{Mitigation strategies for delays in software security patch management}
 
% We identified a group of strategies implemented by the studied teams to minimise the delays. Further investigation enabled us to identify \textit{where} to apply the strategies in the patch management process. Figure \ref{fig:detailstrategies} presents the strategies grouped by the relevant patch management process phase with the number of references for each strategy (in parentheses). 

\begin{figure*}[ht]
  \centering
  \includegraphics[width=1\textwidth]{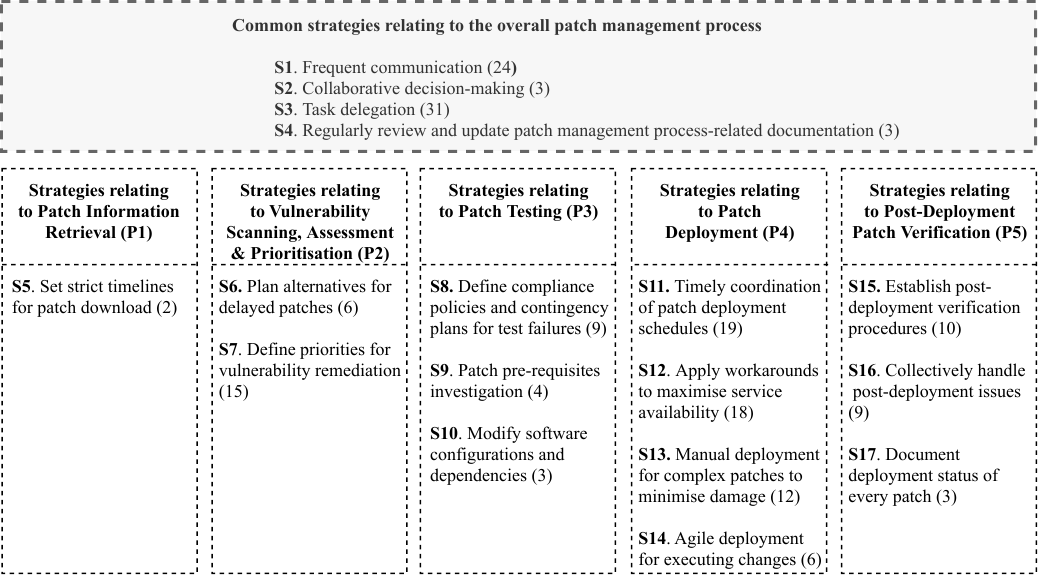}
  \caption{Detailed overview of the strategies applied to mitigate delays in software security patch management.}
  \Description{Strategies for delays in software security patch management.}
  \label{fig:detailstrategies}
\end{figure*}

We identified a group of strategies implemented by the studied teams as corrective/reactive actions to reduce the delays. Further investigation enabled us to identify \textit{where} to apply the strategies in the patch management process. Figure \ref{fig:detailstrategies} presents the strategies grouped by the relevant patch management process phase with the number of references for each strategy (in parentheses). 

\subsubsection{\textbf{Strategies relating to the overall patch management process}.} The following set of \textit{common strategies} can be applied across all phases of the patch management process. \\

\textit{Frequent communication} with all internal and external stakeholders is vital in reducing the patching delays as it helped strengthen the collaboration and improve mutual understanding by bringing all stakeholders on the same page. Regarding internal communication, the studied practitioners held bi-weekly patch meetings to discuss patching issues, find solutions to the issues, report the status of patching tasks, and measure the progress of the patch cycle. Besides the patch meetings, they held informal discussions on complex and critical issues when required.
\begin{quoting}
\textit{\emph{[Subject - Post-deployment issue - Data Capture servers not able to communicate with [system s1]]} \\
``7/8/20 - [P1-AT1] checking with [P2-AT2] for the other three servers that do not have a commissioning request. \\
21/7/20 - Set up another meeting with BT1 to discuss this request (ID 1772737)."} - EMR, Issue ID 42
\end{quoting}

As to external communication, the practitioners frequently negotiated with customers about the patch deployment schedules. It involved getting consent for patch deployment at customers' premises, agreeing on the patch deployment dates and times (i.e., patch window), establishing contact persons at the customer sites for emergency contact and notifying completion of the patch deployment task. Similarly, the practitioners regularly negotiated with the vendors regarding the delayed patch releases and support cases raised for faulty patches. Frequent communication helped all stakeholders gain awareness of the tasks and schedules, assisting them with up-front planning and coordination of the dependent tasks.
\begin{quoting}
\textit{\emph{[Subject - Unix patching schedule confirmation]} \\
``24/7/20 - The requirements analysis revealed a major OS upgrade, not simple patching. The schedule is still being negotiated with [customer c1]."} - Non-Win, Issue ID 7 \\
\end{quoting}

% This way most of the socio-technical reasons causing the delays such as coordination (R3), input-requirements (R4), service availability restrictions (7), and failures from poor planning (R9) appeared to be minimised.\\

% \textbf{S2. Collaborative decision-making.} 
\textit{Collectively making decisions} about patch management, for example, patch prioritisation based on the vulnerability assessment results and organisation needs, selecting workarounds for delayed patching and post-deployment issues helped the team members gain insight into the prospective plans and activities. In addition, it allowed the individual team members to make well-informed decisions about their task assignments that reduced the impact of the delays from waiting for input from dependent tasks and changes in the organisation's schedules.
\begin{quoting}
\textit{\emph{[Subject - Proposal for a patch cycle change in [servers s1 and s2]]} \\
``4/4/18 - Discussions still ongoing for the decision. AT1 is still considering various options and has put them out in slides for discussion at the meeting."} - EMR, Issue ID 2 \\
\end{quoting}

%\textbf{S3. Task delegation.} Task assignment (31) including details of the task, task assignee, raised by, and date of assignment appeared a useful strategy employed by the studied teams. 
We observed the patch meeting facilitator \textit{delegating the tasks} to BT1 team members based on their expertise and experience during the patch meetings. In rare cases, the practitioners voluntarily self-assigned the tasks based on their interests and due to the unavailability of task assignees. The delegated tasks including details of the task, task assignee, raised by, and date of the assignment were documented in the patching tracker during the meeting. It appeared a useful strategy to increase dependency awareness of the tasks, particularly, in scenarios like task B is dependent on task A (A $\rightarrow$ B) and the assignee of task B needs input from task A to progress with the task. 
% In such cases, being aware of the dependent task assignees helps each other to proceed with confidence and minimal impediments. 
Moreover, employing this strategy ensured well-defined roles and responsibilities around patch management activities resulting in increased accountability for actions.
% Consequently, it benefited the studied teams in reducing the delays arising from the lack of coordination (R3), late input requirements (R4), and capacity limitations (R6).
\begin{quoting}
\textit{\emph{[Subject - Vulnerabilities in .NET Core]} \\
``21/2/20 - .NET Core is not receiving updates. A new process is required to patch this version and a service request (SR) needs to be submitted for review and assessment. [P1-BT1] to raise the SR for the issue raised by BT1 on 7th Feb 2020."} - Win, Issue ID 40 \\
\end{quoting}

% \textbf{S4. Regularly update process documentation.} 
Another common strategy that emerged from the data analysis was having a systematic process to \textit{regularly review and update the documentation} about patch management process actions and decisions. It is important to consistently review the process and test any process changes internally before documenting them. A well-documented process ensures clarity in the process activities and decisions and eases tracing back during troubleshooting post-deployment errors. % As a result, employing this strategy helps overcome delays arising from the complexity of patches (R1) and failures from poor execution (R9), particularly relating to inadequate post-deployment patch verification. 
\begin{quoting}
\textit{\emph{[Subject - Update documentation for the split of [servers s1 and s2] patching into two procedures]} \\
``13/12/19 - Finalising the documentation after testing internally for handover to 24x7. \\
10/1/20 - Documentation to be tested in February, will be ready for handover in March."} - EMR, Task ID 24
\end{quoting}

% \subsubsection{Strategies relating to a specific patch management process.} We identified another group of strategies relating to each phase of the patch management process, mentioned in \textit{italic} in the descriptions below. \\

\subsubsection{\textbf{Strategies relating to patch information retrieval (P1).}}

% \textbf{S5. Set strict timelines for patch download.} 
\textit{Setting tight timelines for patch download}, for example, within two days of the \textit{``Patch Tuesday"} when large vendors like Microsoft, Adobe, and others release the patches, was a strategy followed by the studied practitioners. It allowed them sufficient time to plan and coordinate the patch windows, negotiate with customers, obtain organisation approval, and undertake extensive patch testing before deployment. In the studied context, Org B provided a report to Org A teams containing a list of the retrieved patches each month that aided collaborative assessment of vulnerability risks. 
% It was useful to provide advance notice to customers and for in-house application developers' consideration, resulting in reduced delays from the complexity of patches (R1), particularly relating to the increasing rate and diversity of patch release and faulty patches, service availability restrictions (R7), organisation approvals-related reasons (R8), and failures occurring from poor planning (R9), such as missing patch prerequisites due to lack of proper testing and inaccurate estimates in patch patch windows due to rushed planning.  
\begin{quoting}
\textit{\emph{[Subject - Provide .NET report at the start of the patch cycle]} \\
``15/3/19 - Org A requests BT1 to provide an extract of .NET released patches every month and a report including what patches will be applied to what servers."} - EMR, Task ID 53
\end{quoting}

\subsubsection{\textbf{Strategies relating to Vulnerability Scanning, Assessment and Prioritisation (P2).}}

We observed the practitioners \textit{planning alternatives} for scheduled patching that will be delayed due to known reasons. For example, a major upgrade for critical legacy software is a complex and time-consuming process that often involves several challenging subtasks like an intensive assessment of the cost-benefit analysis and impact on other services, and laborious data migration procedures. In such cases, the practitioners planned alternatives (i.e., \textit{what to do} and \textit{when to do it}) for the time being until the software is patched to minimise the risks of attacks. We observed them collaboratively analysing various workarounds for suitability during delayed patch releases and delayed patching and assessing the timing of those alternate remediation plans. \\
% \begin{quoting}
% \textit{\emph{[Subject - Org A Change Freeze from 14th December to 12th January]} \\
% ``31/10/19 - Due to the change freeze in December, there will be no patching for December. \\
% 1/11/19 - AT2 wants to reserve the first two weeks of December for remediation and out-of-band patching of critical vulnerabilities as required."} - Win, Task ID 34 \\
% \end{quoting}

\textit{Defining priorities for vulnerability remediation} appeared beneficial in reducing the risk of exploitable attack vectors from delayed remediation due to the large number and diversity of patch releases. The studied practitioners prioritised vulnerabilities based on the patch severity and impact. In the studied context, the security team (AT4) prioritised security patches based on the global vulnerability rating and their own risk assessment. High-risk critical patches were prioritised to be deployed within 48 hours while the medium to low-risk patches were deployed in the next patching cycle. Prioritisation based on the patch type, for example, operating system patch vs software application patch, was another strategy employed for defining the priorities. In some cases, we observed them prioritising the operating system security patches over other security patches like .NET, IE, Adobe, and Java.
\begin{quoting}
\textit{\emph{[Subject - OS security patches need to be tracked separately in the vulnerability remediation]} \\
``15/5/20 - [P1-AT1] requesting the OS security patches to be tracked separately from all other vulnerability remediation. Org B's report should only be addressing OS security patches anyway but can make sure to separate any non-OS remediation tasks."} - EMR, Task ID 45
\end{quoting}

\subsubsection{\textbf{Strategies relating to Patch Testing (P3).}}

\textit{Definition of compliance policies}, for example, the standards imposed by the security team to reboot every legacy server even if there are no patches, and developing contingency plans in cases of failures appeared beneficial in mitigating the risk of delays caused by the erroneous patches. \\
% \begin{quoting}
% \textit{\emph{[Subject - [s1] new servers compliance]} \\
% ``4/12/18 - This item remains open until all new [s1] servers are fully compliant including security hardening prior to being pushed into production and support. \\
% 18/2/18 - Security approved the new [s1] servers, go-live completed on 18/2."} - EMR, Task ID 1 \\
% \end{quoting}

Patch prerequisites such as the registry changes and preparation package installation represent preconditions that needed to be set up for the patch to take effect during the deployment. As a strategy to avoid possible delays resulting from the runtime errors hindering patch deployment due to missing prerequisites and delays in manual configurations associated with the prerequisites, the BT1 team performed an \textit{investigation of prerequisites} for the patches released every month as a separate task during patch testing. 
% It was discussed in the patch meetings each month and the progress was tracked in the patching tracker. 
\begin{quoting}
\textit{\emph{[Subject - Registry key missing for Knowledge Base (KB) ID [n] (LDAP)]} \\
``2/10/20 - Patches not installed on [servers s1 and s2] due to missing a registry key. [P1-BT1] to check settings and apply where missing."} - Win, Task ID 24 \\
\end{quoting}

In preparing the machines for patch deployment and avoiding potential delays arising from complexities of patches due to patch dependencies, the practitioners dedicated a specific time to \textit{identifying and modifying the dependencies and configurations} during patch testing. For example, they created patch clusters based on the patch similarity and configured the group settings, also known as Group Policy Object (GPO), to reduce time spent on manual configurations on individual patches. 
% \begin{quoting}
% \textit{\emph{[Subject - GPO creations (configs) to be done together]} \\
% ``12/6/20 - [P1-BT1] to create new GPOs at the same time early next week in preparation for the next round of additions for vulnerability remediation."} - Win, Task ID 49 \\
% \end{quoting}

\subsubsection{\textbf{Strategies relating to Patch Deployment (P4).}}

\textit{Well-timed coordination of patch deployment schedules} can help mitigate several delays associated with the coordination delays, capacity limitations, organisation policies regarding service availability, organisation schedule changes, failures from poor planning, and increased rate of patch release during patch deployment. The activities involved internal planning and scheduling of the patch windows for each managed system (i.e., \textit{when to patch}), defining the teams' roles and responsibilities for contacting customer sites for patch deployment verification and planning the servers' load to spread evenly through the patch windows to avoid performance issues and unexpected service disruptions during patch deployment (i.e., \textit{how to patch}). \\
% \begin{change}
% \begin{quoting}
% \textit{\emph{[Subject - Review of [s1] servers' patch windows: re-balancing and extended windows proposed]} \\
% ``13/3/19 - Org B proposes 4-hour windows starting at 18:00 each night. The first lot of servers to start Friday week 2 after the "Patch Tuesday". Sample of re-balanced servers provided for Org A's consideration."} - Win, Task ID 9 \\
% \end{quoting}
% \end{change}

% \textbf{S12. Maximise availability by applying the workarounds.} 
Given the mission-critical nature of healthcare operations, the risk of system downtime from reboots presented a major challenge to the practitioners in reducing the risk of service disruptions during patch deployment. As a strategy to maximise service availability and reduce potential associated delays, they \textit{applied various countermeasures including clustering, load balancing, and failover}. Clustering refers to grouping patches based on their similarity. As such, configuring the group settings and deployment of the patch clusters significantly reduced the time spent in testing, deployment, and rebooting than comparable single-patch work resulting in increased service availability. Similarly, load balancing which refers to balancing the load on servers during deployment helped avoid unnecessary service disruptions. This is because the servers will be patched in batches reducing the risk of all services being interrupted at the same time. Failover or maintaining backup servers to concurrently run the services while being rebooted was another workaround employed to minimise the downtime. Subsequently, the backup servers' patching was carefully planned with separate patch windows. A few other countermeasures included planning extended windows for patches that required multi reboots in out-of-band windows and pre-loading the patches offline to avoid patch deployment exceeding the allocated patch window.
\begin{quoting}
\textit{\emph{[Subject - Patch deployment failed at [server s1]]} \\
``24/1/18 - Single point of failure for [server s1]. AT1 to review the proposed design for clustering for high availability. Currently hard to obtain reboot timings and only one reboot is allowed. Ask the customer for an extended window and move the patching to the weekend."} - Win, Task ID 6 \\
\end{quoting}

% \textbf{S13. Minimise damage by shifting to manual deployment for complex and faulty patches.} 
The practitioners decided to \textit{shift to manual patch deployment} for business-critical server patching, complex patches that involved multiple version dependencies, multi reboots and legacy software systems, and redeployment of erroneous patches. This strategy was deemed effective in minimising the damage (i.e., service operations left unstable post-deployment) caused by failed deployments and avoiding the risk of further delays. % Furthermore, as a consequence, it allowed the patching delays from service availability restrictions (R7) and complexity of patches (R1) relating to patch interdependencies and faulty patches to be minimised. 
However, we noticed that shifting to the manual deployment itself could lead to delays in patching as described in R3 in Section 4.1.1. \\
% \begin{quoting}
% \textit{\emph{[Subject - Post-deployment error at [server s1] causing the printing service unavailable]} \\
% ``31/10/19 - This server was patched on 18/10 at 2 am-6 am. Due to the errors, patching will be done manually in November."} - Win, Issue ID 6 \\
% \end{quoting}

% \textbf{S14. Agile deployment by executing complex patch deployment changes in small iterations.} 
\textit{Agile deployment} was another strategy employed by the teams where they executed the changes to patch deployment procedures in small iterations. This was adopted as a precautionary measure against unexpected breakdowns since a small change in the deployment process could result in disastrous consequences to service continuity and build confidence around the new changes.
% Consequently, it seemed beneficial in reducing the delays associated with service availability restrictions (R7) and extensive monitoring needed for verification of impact from large changes (R1).
% \begin{quoting}
% \textit{\emph{[Subject - Review of patch cycle timings]} \\
% ``22/2/18 - Org A is considering going for bi-monthly deployment cycles for .NET patching and then move to monthly deployment after the confidence is built."} - EMR, Task ID 2 
% \end{quoting}

\subsubsection{\textbf{Strategies relating to Post-Deployment Patch Verification (P5).}}

% \textbf{S15. Establish post-deployment verification procedures.} 
\textit{Having a defined set of procedures for post-deployment patch verification} helps reduce the risk of delays caused by failures from poor execution due to inadequate post-deployment verification. The studied teams verified the patch deployment status using several approaches such as monitoring the system for any functional, performance, or unexpected issues, analysing the system logs, collecting user feedback (i.e., confirming with customers about any adverse impact on service continuity), and getting periodic scans to verify the targeted security vulnerabilities have been patched.
\begin{quoting}
\textit{\emph{[Subject - Automated second rescan for reboots]} \\
``31/10/19 - [P1-BT1] raised this issue, he has configured the window to rescan for missing patches and conduct a second reboot if required. No issues during patching, seeking client feedback for verification."} - EMR, Task ID 28 \\
\end{quoting}

% \textbf{S16. Collectively handle post-deployment issues.} 
Post-deployment issues such as unresponsive server or unavailability of service may have developed due to failures during patch testing and deployment, or lack of proper post-deployment verification. To avoid such issues leading to long delays causing unexpected service disruptions, the practitioners engaged in a \textit{collaborative problem handling approach}. We observed long discussions at the patch meetings about the investigations of the root causes for post-deployment issues and finding workarounds to failed deployments. Most commonly used workarounds in the studied context included reverting to the previous working software version, restoring from the backup, and patch redeployment in out-of-band windows. \\
% \begin{quoting}
% \textit{\emph{[Subject - Tracking of GPO applications that are not intended]} \\
% ``10/7/20 - Information sent to [P1-AT2] and [P2-AT2]. [Patch p1] will need a rollback. \\
% 24/7/20 - Rollback executed, no reported issue due to the rollback. Keep open for one more meeting for monitoring."} - Win, Task ID 51 \\
% \end{quoting}

% \textbf{S17. Document deployment status of every patch.} 
The team members \textit{documented the deployment status of every patch} in the patching tracker. It served useful as a vulnerability wiki to keep track of the progress of every patch and as a reference in cases of errors encountered during the execution. Further, employing this strategy during post-deployment patch verification ensured all patches are properly deployed and audited. As a result, the delays that occurred due to tool limitations, for example, missing patches during deployment were minimised. \\ 
% \begin{quoting}
% \textit{\emph{[Subject - [Servers s1 and s2] successfully patched]} \\
% ``24/7/20 - [...] No further issues experienced since patching. Manual instructions and deployment status updated in the shared tracker. Will be kept in-monitor for another couple of weeks."} - EMR, Issue ID 47
% \end{quoting}

% \noindent\fbox{%
%     \parbox{\textwidth}{%
%         \textbf{Summary of RQ2:} The data analysis revealed a collection of strategies applied by the studied practitioners to minimise patching delays. We presented 17 strategies grouped by the patch management process phase, describing how they helped address the delays.
%     }%
% }

\noindent\fbox{%
    \parbox{\textwidth}{%
        \textbf{Summary for RQ2:} We identified 17 strategies applied by the practitioners as corrective/reactive actions to manage the delays. Among these strategies, frequent communication, collaborative decision-making, task delegation, and regularly reviewing and updating the documentation were common strategies applied across all phases of the security patch management process. Further, we found a group of strategies executed at each phase of the process to mitigate the delays that occurred during each step.
    }%
}
% \noindent\fbox{%
%     \parbox{\textwidth}{%
%         \textcolor{mediumblue}{\textbf{Key Findings}:\\
%         • Identified 17 strategies applied by the practitioners as corrective/reactive actions to manage the delays. \\
%         • Among these, four common strategies are applied throughout the security patch management process. \\
%         • The other 13 strategies are distinctly applied to the five phases of the process.}
%     }%
% }

\subsection{Findings from Member checking}

The participants provided positive feedback on the study findings and agreed with the accuracy of the results. % No new information or changes to the current findings emerged from the participants. 
Several participants including the executive complimented our research, saying \textit{``Thanks for all the information. Very interesting analysis"}- P1-Org A, \textit{``From my point of view, I think your analysis is very good and useful because it's not just looking at how good or bad things are but also highlights where the improvement could be"}- Executive-Org A. Further, it was interesting to see their motivation to improve the delays following the presentation. \textit{``I hate to see this good work going wasted, a really good analysis where we got some really good insights. So, I'd like to see our teams taking these on board, then revisit this to see how the pie chart changes when we address the top reasons for delays"}- Executive-Org A. % For some reasons for delays, the participants explained the challenges encountered while handling them,
The participants did not mention any new information or variations to the findings and explained the challenges of dealing with some of the delays, for example, \textit{``The patching timeline is fixed by vendors such as Microsoft who use a monthly schedule so reducing the time frame of getting appropriate approvals and executing is an absolute necessity. And getting new patches tested, confirmed, and approved in a week is always a challenge before they are rolled out confidently to production"}- P1-Org A, \textit{``Also, not all environments have testing environments to test these patches. So, in a fair few cases application testing actually occurs in deployment environments which can cause many failures leading to delays"}- P2-Org B, \textit{``Yes, to add to it, vendors introducing application patches at the same time as OS patches can also cause delays and conflicts with OS security patching"}- P3-Org B. They also asked us several questions including how they can reduce the delays further, to which we suggested some improvements which are discussed in Section \ref{section:discussion}.

\section{Discussion}
\label{section:discussion}

In this section, we reflect upon our findings and discuss them in light of the existing literature. Further, we present the implications for research and practice. \\
% \subsection{Key findings}

Mitigating delays in security patch management is instrumental in maintaining the security, availability, and confidentiality of information technology (IT) systems \cite{nist2005guide}, and failure to do so has resulted in several devastating outcomes \cite{Equifaxbug}. Yet, the topic remains less explored in the literature, particularly, in understanding the practical reasons for delays in applying the patches. Based on a comprehensive analysis of the gathered artefacts over a period of four years, we have identified why, how and where delays happen in security patch management in practice and a set of corrective strategies to mitigate them.

Our findings unveil that the primary cause of the most prevalent delays (24.9\%) is coordination delays in the patch management process (Figure \ref{fig:delaysfrequency}). The need for effective coordination in patch management appears from a combination of complex technical dependencies inherent in patches and the collaborative environment in the patch management process demanding orchestrated social dependencies between a diverse group of stakeholders \cite{dissanayake2021grounded}. An interesting finding is that internal coordination delays were more recurrent than the delays occurring from external stakeholders, i.e., customers and vendors. This was confirmed during the member checking as described by the executive, \textit{``I'm not surprised by some of these reasons, especially the coordination delays as the difficulties in collaborating and communicating between the teams are evident in almost every aspect of the process."} Although it appears that such delays are within the control of the practitioners, our findings emphasize the need for further support on coordination across patch management tasks and stakeholders. Similarly, with regards to the second most recurrent reason, the input requirements delays (16.8\%), a majority of the delays emerged from internal teams as opposed to external vendors, indicating that adopting strategies like frequent communication (S1) and task delegation at meetings (S3) can help reduce such delays. 

The next prominent reason, the complexity of patches (13.7\%) can be attributed to the inherent complex patch dependencies and unknown risks of faulty patches. Although the intrinsic factors are essentially in control of the third-party vendors in charge of patch development, strategies like extensive patch testing to identify the prerequisites and inherent patch dependencies (S9, S10) \cite{li2019keepers, dissanayake2021grounded, dissanayake2021software} and defining contingency plans to handle faulty patch errors (S8) can help reduce delays arising from the patch complexity. The socio-technical endeavour in patch management constituting the fourth-most recurring delay (13.4\%) can be explained by the inevitable need for human intervention in the process. While it suggests a need for a better understanding of the human interaction in patch management, our findings can guide practitioners in the planning of patch schedules allowing sufficient time for manual intervention (S11). Regarding the delays caused by capacity limitations in human resources, infrastructure, and time (9.6\%), properly planning the task assignments with minimum task dependencies (S3), patch clustering and load balancing (S12), and implementing patch deployment changes in an agile manner (S14) can be helpful.

While organisation-related delays (7.4\%) can be implied to be within the control of practitioners, service availability restrictions (6\%) may appear difficult to always be taken control of. This is because service continuity presents a pressing need for modern enterprises, particularly in the context of mission-critical domains for which service disruptions even for a few seconds can result in severe consequences. As described by a participant during member checking, \textit{``it is very challenging with the service availability restrictions, one example is the ambulance service, even though we have received approval, we always have to call the service just to confirm if it's okay to patch because we don't want to shut down the system in the middle of an operation"}. However, applying workarounds such as failover, clustering, and load balancing (S12) can help reduce such delays. Concerning the least occurring delays, limitations of existing tools (4.6\%), although reflect reasons not within practitioners' control, having well-established roles, patch management practices, and policies can help mitigate such delays. Finally, the delays emerging from failures in poor planning and execution (3.6\%) can be addressed with careful planning and execution (S9-11, S17). \\

% Further reflecting upon our findings and in comparison to previous works, we identify that some reasons are \textit{general} and could be relevant to other domains too while some of them are likely to be \textit{specific} to the cases studied based on their characteristics.

% Regarding \textit{general} reasons, we identify that these reasons include a set of factors that are integral to security patch management. Thus, they are recognised as standard reasons for delays that could be generalizable beyond the cases studied. 
Further reflecting upon our findings and in comparison to previous works, we discuss that some of the identified reasons are not necessarily specific to security patching in the domain of healthcare, but could be also observed in other domains. For example, the \textbf{complexity of patches (R1)}. The patch interdependencies are found to be intrinsic characteristics present in the patches released by the vendors \cite{dissanayake2021grounded}. Therefore, the resultant delays from managing these patch interdependencies could be challenging in other domains as well. Similarly, the need of human expertise is a standard notion accepted in security patch management because the process is inherently a socio-technical endeavour, where the human and technical interactions are tightly interconnected \cite{nicastro2003security, dissanayake2021software}. Therefore, we find the reasons relating to the \textbf{need of human intervention (R3)} as reasons that could also apply beyond the studied context. In addition, we recognise the reasons relating to the \textbf{service availability restrictions (R9)} could be present in other domains as well. This is because the reboots following patch deployment are necessary for the applied patch to take effect. Further, the service interruptions caused by the reboots have been widely acknowledged as a major obstacle in patch management across several domains \cite{potter2005reducing, dumitracs2009upgrades, li2019keepers, tiefenau2020security, araujo2020improving}.

% In contrast, we identify another group of reasons that could be \textit{specific} to the cases studied (i.e., organisation and domain bound). We classify them as context-specific reasons as they include a set of contextual factors. Examples of context-specific reasons are \textbf{organisation delays (R7)} and \textbf{capacity limitations (R8)}. Concerning organisation-related reasons, patching delays resulting from delayed approval from higher management may not directly apply to a small organisation with one team or to an organisation with a flat hierarchy where no line approvals are needed. Although these reasons may not necessarily represent reasons beyond the studied case or the context, an understanding of the context-specific factors enables researchers and practitioners to better appreciate the practical utility of the solutions and formulate appropriate plans for mitigating potential delays. Moreover, as mentioned in section 5.3, there are possibilities for future research to explore the reasons for delays in a broader context using these categories.

In contrast, we believe that some of the reasons are likely to be specific to the domain of healthcare and the context of studied organisations. For example, the reasons attributing to \textbf{organisation delays (R7)} and \textbf{capacity limitations (R8)}. Concerning organisation-related reasons, patching delays resulting from delayed approval from higher management may not directly apply to a small organisation with one team or to an organisation with a flat hierarchy where no line approvals are needed. Although these reasons may not necessarily represent reasons beyond the studied cases or the context, an understanding of the context-specific reasons enables researchers and practitioners to better appreciate the practical utility of the solutions and formulate appropriate plans for mitigating potential delays. We believe there are possibilities for future research to explore the reasons for delays in a broader context using these categories.

\subsection{Related Works}

Our study confirms the findings of the previous studies that suggest some challenges in security patch management could contribute to delays in patching. For example, our finding of coordination delays contributing to the majority of the delays complements the existing research \cite{nicastro2003security, nappa2015attack, li2019keepers, huang2012patch, potter2005reducing}, which reported that coordination is one of the most pressing challenges of timely patch management. Our analysis extends the knowledge by showing how coordination delays are introduced internally and externally. Additionally, our findings further highlight the importance and the need to focus more on the socio-technical aspects such as coordination in the time-critical security patch management process as mentioned by previous literature \cite{nicastro2003security, li2019keepers, tiefenau2020security, dissanayake2021software, dissanayake2021grounded}. 

Our analysis reveals that the complexity of patches causes the third-most frequent reason for delays; it complements the previous work \cite{nappa2015attack, dunagan2004towards, tiefenau2020security, jenkins2020anyone, crameri2007staged, nicastro2003security, tucek2009efficient}, which has mentioned that faulty patches and configuring patch dependencies are challenging as they often lead to breakdowns during patch deployment. Similarly, the need of human expertise in the process \cite{li2019keepers, tiefenau2020security, crameri2007staged, dissanayake2021grounded} and capacity limitations, specifically, lack of human resources \cite{post2003computer, tiefenau2020security, jenkins2020anyone} are mentioned as challenges in security patch management in the related studies. Further, several studies (e.g., \cite{potter2005reducing, dumitracs2009upgrades, li2019keepers, tiefenau2020security, araujo2020improving}) have highlighted service disruption as a central challenge of patch deployment. Our study extends the knowledge of these challenges by showing how, why and when they contribute towards patching delays.

Alternatively, previous studies \cite{cavusoglu2008security, cavusoglu2006economics} have predominantly focused on achieving timely patch management through optimising the process by attaining a balance between an organisation’s patch cycle and a vendor’s patch release cycle. Dey et al. \cite{dey2015optimal} have developed a quantitative framework that analyses and compares various patching policies to find the optimum policy considering the costs of periodic patching against the security risks from patching delays. By investigating vendors' patch release and practitioners' patch deployment practices, Nappa et al. \cite{nappa2015attack} revealed that only 14\% of the patches are deployed on time and the patching mechanism (e.g., automated vs manual patch deployment) impacts the rate of patch deployment. Despite the widespread attention towards timely security patch management, an important observation is the absence of an investigation of the root causes (i.e., reasons) for delays in security patch management. To the best of our knowledge, the existing studies have not explored why the application of patches is delayed but rather proposed approaches to achieve a timely patch management process. Hence, our study contributes to the existing body of knowledge by: \\
  - providing a taxonomy of reasons explaining why delays occur when applying security patches in practice, \\
  - reporting what reasons for delays are more prominent based on frequency analysis, \\
  - demonstrating where the delays occur in the patch management process, \\
  - presenting a set of strategies to mitigate the delays and describing when they can be applied in the patching process, \\
  - providing practical implications for practitioners to identify and mitigate delays, and, \\
  - establishing a foundation for future research towards effective management of patching delays.

\subsection{Implications for Practitioners}

Our findings reveal why delays happen when applying security patches in practice with a set of reasons contributing to the delays, explain how the reasons vary, and how delays are distributed in the patch management process. As a direct practical implication of the provided understanding, the security analysts and system administrators will be able to identify and assess the factors associated with the causes of delays and take precautions to mitigate potential delays. Further, the understanding of the frequency analysis of reasons and distribution of the delays highlights \textit{what} reasons need practitioners' immediate attention and \textit{where} is improvement needed to overcome the delays. % suggest the importance of recognising such socio-technical gestures in patch management, which are often overlooked in practice. 
In addition, the knowledge will help practitioners in suitable decision-making, prioritisation, and planning of patch management tasks with minimal impediments. 

In addition to explaining why, how, and where do delays occur in patching, our findings describe how the delays can be mitigated. We present a set of strategies employed by the studied practitioners to rectify the delays. Knowing what to do and when to do can be useful for practitioners and organisations in taking prompt actions to mitigate the impact of the delays. The findings may also help predict a delay in a given scenario whereby practitioners can better plan patch cycles and refine the patching process in light of their organisational contexts. For example, practitioners can consider the development of new tools like Environment Diagrams as a visualisation tool, to keep track of the system dependencies that would save time in patch testing and deployment. Other approaches like maintaining an online shared repository documenting organisation schedules and regularly documenting patch exemptions in detail would assist teams with accurate planning of patch schedules. Towards overcoming delays of coordination in patching, adopting computer-supported collaborative tools like ``Slack" can benefit accomplishing timely communication, collaboration, and information sharing between all stakeholders \cite{lin2016developers}. In this way, our findings offer guidance to practitioners to make suitable decisions to alleviate the threat of cyberattacks from delayed patching. \\

\subsection{Implications for Researchers}

Given our findings are based on the cases studied limiting to the domain of healthcare, other researchers can extend and adapt the results through future studies within the same domain involving different stakeholders or different domains. Further, future research exploring the viability of the findings based on the contextual factors, for example, variations in context-specific reasons for delays like capacity limitations (R8) and organisation delays (R7), can result in useful insights from additional cases with extended scope. With regards to the reported strategies for mitigating the delays, future studies can investigate their suitability and effectiveness depending on the context and organisation policies (e.g., similar to future work of \cite{li2019keepers, tiefenau2020security, dissanayake2021grounded}). In addition, the findings can be used in potential interview guides and surveys to verify the findings in other contexts and discover variations within them. Another possibility is to investigate the impact of patching delays on organisations and other stakeholders such as end-users.

The data analysis has revealed that the limitations in current tools contribute to delays in applying the patches. We believe that future research can address this limitation by developing advanced tools leveraging deep learning techniques. For example, an automated tool that provides dependency visibility by highlighting mismatches of patch dependencies. Considering the most recurrent delays occur due to a lack of coordination in the patching process and delays in providing the required inputs, future research can invest efforts into developing computer-supported tools and platforms that can support better coordination across patching tasks and reducing delays in collaborative tasks. Solutions could be further investigated on how automation support can be extended to assist the decision-making in patch management, for example, developing intelligent interactive systems like software bots \cite{10.1145/3274451} for collaborating with practitioners that guide them to decisions by asking rational questions. This further opens up an avenue for future research to explore how \textit{``human-AI collaboration"} \cite{kamar2016directions}, an emerging research paradigm in the CSCW community \cite{wang2019human}, can be extended to a crucial topic like security patch management. Moreover, there is room for research to explore how to improve the performance and accuracy of the patch management tools. Tool development needs to consider the diversity of operating systems, software applications, platforms, and programming languages in vulnerability patch management to overcome the obstacles of lack of accuracy and scalability in the current tools (R2). In addition, the researchers, particularly the usable security researchers, can study how to improve the design of such smart tools. \\

\section{Threats to Validity}
\label{section:threatstovalidity}

In this section, we discuss the potential threats to validity and how they were mitigated following the guidelines proposed by \cite{runeson2009guidelines, yin1994case, maxwell1992understanding}. \\

\textbf{External validity - Generalizability: } % A Grounded Theory does not generalize to a population as the theory is developed limited to the studied context or cases \cite{strauss1998basics, glaser1978theoretical, glaser1967discovery}. The data collected is limited to the studied cases in security patch management in the healthcare domain. Hence, our theory about delays in software security patch management may not necessarily represent the context of other domains. However, we believe that our findings can be recreated and adapted in other contexts.
This study is based on the empirical data collected from a particular context, i.e., security patch management in the healthcare domain. Hence, our findings do not claim for generalization to all other contexts of patch management, instead, this study focuses on performing a comprehensive investigation of the delays in security patch management within the studied setting to provide detailed explanations through rigorous data analysis. However, we do not assert the results to be absolute or final, rather they can be recreated and adapted in other contexts \cite{strauss1998basics, glaser1978theoretical, glaser1967discovery}.

Regarding data representativeness, the study includes data collected limited to the patching tracker. However, collecting data from two organisations with multiple teams including participants with diverse roles and wide experience increased the data reliability and assisted in ensuring participant triangulation \cite{maxwell1992understanding}. Although we have analysed data spanning over four years from October 2016 to May 2021, it is possible that we may have missed some variations in the findings, specifically the context-specific reasons and strategies. We suggest that any future studies on this topic include more data sources such as additional cases or interviews to extend the scope of our findings and verify their explanatory power in other contexts. 

\textbf{Reliability: } To mitigate the threat of subjectivity % integral to GT procedures 
and ensure reliability in the data analysis, all the data collection and analysis procedures, emerged codes, and identified relationships were discussed in detail among all authors and finalised through multiple revisions. In addition, related to interpretive validity \cite{maxwell1992understanding}, we conducted member checking to verify the accuracy of our findings, which was attended by three authors, further ensuring investigator triangulation.  
% Further, emerging at the final theory through multiple revisions of models from the coding, which was regularly discussed over several months by the co-authors ensured application of Theory triangulation.

\textbf{Construct Validity: } To address the threat of construct validity, we used multiple sources of evidence, i.e., analysis of artefacts and observations, and multiple stakeholders, maintained a chain of evidence (e.g., the coding procedure following the Grounded Theory method \cite{rodriguez2020theory}), and had the findings reviewed through member checking.

\textbf{Internal Validity: } To mitigate the threat of internal validity and misrepresentation, we ensured participant triangulation by covering the entire population involved in security patch management representing all teams in both organisations. In addition, the participants had a wide experience in security patch management, which helps mitigate the risk of participants' lack of expertise.

\textbf{Evaluative Validity: } The verifiability of the findings that emerged from a grounded theory data analysis can be attained from the adequacy and soundness of the research methodology through which the findings emerge \cite{strauss1998basics, glaser1967discovery}. To achieve this, we have detailed our data analysis process of the application of the Straussian GT procedures in Section \ref{section:researchmethod}. Further, to alleviate the reporting bias, we have included quotes from the patching tracker in Section \ref{section:findings}. % Hence, these details provide evidence of how our theory fulfills the evaluation criteria proposed by Strauss \cite{strauss1998basics} about the validity, reliability, and credibility of the data, theory, and theory development. 

\section{Conclusion}
\label{section:conclusion}
% Applying timely software patches to identified security vulnerabilities is crucial to system security. Failing to patch software security vulnerabilities on time has resulted in several devastating consequences including human death. While it is important to understand why practitioners fail to apply security patches on time and explore possible mechanisms to reduce such delays, there is little empirical evidence to explain why it continues in practice. This study attempts to answer these overarching questions by explaining why, how, and when patching delays occur in practice, and what strategies can be applied to mitigate delays. Through a longitudinal study representing eight different teams from two organisations in the domain of healthcare, and based on a Grounded Theory analysis of 132 delayed tasks documented in the patching tracker over a period of four years from October 2016 to May 2021, we identify a set of technical and socio-technical reasons that cause delays when applying the security patches. Towards reducing such delays, practitioners apply several strategies based on the phase of the patch management process the delays occur. Further, we provide insights into how the delays are distributed in the patch management process and the frequency of occurrence of reasons for delays, highlighting \textit{where} and \textit{what} in the patch management process needs urgent attention from practitioners.

In this study, we empirically explore and systemically explain why, how, and where delays occur when applying security patches in practice, and how the delays can be mitigated. Through a longitudinal study representing eight different teams from two organisations in the domain of healthcare, and based on a Grounded Theory data analysis of 132 delayed tasks documented in the patching tracker over a period of four years from October 2016 to May 2021, we identify a set of reasons relating to technology, people and organisation that cause delays in security patch management. We also provide an evidence-based understanding of the frequency distribution of reasons for delays and distribution of delays over the patch management process. Such information highlights the reasons that need immediate attention and the areas of improvement in the patch management process. Additionally, we report a set of strategies that can be used for mitigating the delays in applying security patches by practitioners.

Compared to the related literature, our study provides a holistic understanding of the delays when applying security patches in practice; it is the first attempt to empirically investigate the topic in-depth. We assert that the reported understanding of why, how, and where delays occur during patching and how they can be mitigated will help practitioners take suitable decisions to mitigate delays and guide them towards taking timely actions to avoid potentially disastrous consequences from delays in patching. Furthermore, our findings lay the foundation for future research to investigate and develop computer-supported tools that can address the practical concerns causing delays in patch management, drawing attention to a topic, critical and timely, yet less explored in the CSCW community.

\begin{acks}
The authors sincerely thank the industry collaborators of CREST, without whose support this research would not have been possible. Thank you for allowing us to participate in your process of security patch management, and for providing us with great feedback on our presentation. We also thank the reviewers for their valuable insights and feedback.
\end{acks}

\bibliographystyle{ACM-Reference-Format}
\bibliography{bibliography}


\begin{thebibliography}{57}

%%% the Web address.

\ifx \showCODEN    \undefined \def \showCODEN     #1{\unskip}     \fi
\ifx \showDOI      \undefined \def \showDOI       #1{#1}\fi
\ifx \showISBNx    \undefined \def \showISBNx     #1{\unskip}     \fi
\ifx \showISBNxiii \undefined \def \showISBNxiii  #1{\unskip}     \fi
\ifx \showISSN     \undefined \def \showISSN      #1{\unskip}     \fi
\ifx \showLCCN     \undefined \def \showLCCN      #1{\unskip}     \fi
\ifx \shownote     \undefined \def \shownote      #1{#1}          \fi
\ifx \showarticletitle \undefined \def \showarticletitle #1{#1}   \fi
\ifx \showURL      \undefined \def \showURL       {\relax}        \fi
% invisible to TeX
\providecommand\bibfield[2]{#2}
\providecommand\bibinfo[2]{#2}
\providecommand\natexlab[1]{#1}
\providecommand\showeprint[2][]{arXiv:#2}

\bibitem[\protect\citeauthoryear{Araujo and Taylor}{Araujo and Taylor}{2020}]%
        {araujo2020improving}
\bibfield{author}{\bibinfo{person}{Frederico Araujo} {and}
  \bibinfo{person}{Teryl Taylor}.} \bibinfo{year}{2020}\natexlab{}.
\newblock \showarticletitle{Improving Cybersecurity Hygiene through JIT
  Patching}. In \bibinfo{booktitle}{\emph{Proceedings of the 28th ACM Joint
  European Software Engineering Conference and Symposium on the Foundations of
  Software Engineering (ESEC/FSE ’20)}}. \bibinfo{publisher}{ACM},
  \bibinfo{pages}{1421--1432}.
\newblock
\urldef\tempurl%
\url{https://doi.org/10.1145/3368089.3417056}
\showDOI{\tempurl}


\bibitem[\protect\citeauthoryear{Birt, Scott, Cavers, and Walter}{Birt
  et~al\mbox{.}}{2016}]%
        {birt2016member}
\bibfield{author}{\bibinfo{person}{Linda Birt}, \bibinfo{person}{Suzanne
  Scott}, \bibinfo{person}{Debbie Cavers}, {and} \bibinfo{person}{Christine
  Campbell~Fiona Walter}.} \bibinfo{year}{2016}\natexlab{}.
\newblock \showarticletitle{Member Checking: A Tool to Enhance Trustworthiness
  or Merely a Nod to Validation?}
\newblock \bibinfo{journal}{\emph{Qualitative Health Research}}
  \bibinfo{volume}{26}, \bibinfo{number}{13} (\bibinfo{year}{2016}),
  \bibinfo{pages}{1802--1811}.
\newblock
\urldef\tempurl%
\url{https://doi.org/10.1177/1049732316654870}
\showURL{%
\tempurl}


\bibitem[\protect\citeauthoryear{Brooks}{Brooks}{1975}]%
        {brooks1975essays}
\bibfield{author}{\bibinfo{person}{Frederick~P. Brooks}.}
  \bibinfo{year}{1975}\natexlab{}.
\newblock \bibinfo{booktitle}{\emph{The Mythical Man-Month: Essays on Software
  Engineering}}.
\newblock \bibinfo{publisher}{Adisson-Wesley}, \bibinfo{address}{London}.
\newblock


\bibitem[\protect\citeauthoryear{Cavusoglu, Cavusoglu, and Zhang}{Cavusoglu
  et~al\mbox{.}}{2006}]%
        {cavusoglu2006economics}
\bibfield{author}{\bibinfo{person}{Huseyin Cavusoglu}, \bibinfo{person}{Hasan
  Cavusoglu}, {and} \bibinfo{person}{Jun Zhang}.}
  \bibinfo{year}{2006}\natexlab{}.
\newblock \showarticletitle{Economics of Security Patch Management}. In
  \bibinfo{booktitle}{\emph{WEIS}}. \bibinfo{publisher}{Citeseer},
  \bibinfo{pages}{1--10}.
\newblock


\bibitem[\protect\citeauthoryear{Cavusoglu, Cavusoglu, and Zhang}{Cavusoglu
  et~al\mbox{.}}{2008}]%
        {cavusoglu2008security}
\bibfield{author}{\bibinfo{person}{Hasan Cavusoglu}, \bibinfo{person}{Huseyin
  Cavusoglu}, {and} \bibinfo{person}{Jun Zhang}.}
  \bibinfo{year}{2008}\natexlab{}.
\newblock \showarticletitle{Security Patch Management: Share the Burden or
  Share the Damage?}
\newblock \bibinfo{journal}{\emph{Management Science}} \bibinfo{volume}{54},
  \bibinfo{number}{4} (\bibinfo{year}{2008}), \bibinfo{pages}{657--670}.
\newblock
\urldef\tempurl%
\url{https://doi.org/10.1287/mnsc.1070.0794}
\showURL{%
\tempurl}


\bibitem[\protect\citeauthoryear{Charmaz}{Charmaz}{2006}]%
        {charmaz2006constructing}
\bibfield{author}{\bibinfo{person}{Kathy Charmaz}.}
  \bibinfo{year}{2006}\natexlab{}.
\newblock \bibinfo{booktitle}{\emph{Constructing Grounded Theory: A Practical
  Guide through Qualitative Analysis}}.
\newblock \bibinfo{publisher}{Sage}.
\newblock


\bibitem[\protect\citeauthoryear{Chen, Kim, Wang, Zeldovich, and Kaashoek}{Chen
  et~al\mbox{.}}{2014}]%
        {chen2014identifying}
\bibfield{author}{\bibinfo{person}{Haogang Chen}, \bibinfo{person}{Taesoo Kim},
  \bibinfo{person}{Xi Wang}, \bibinfo{person}{Nickolai Zeldovich}, {and}
  \bibinfo{person}{M.~Frans Kaashoek}.} \bibinfo{year}{2014}\natexlab{}.
\newblock \showarticletitle{Identifying Information Disclosure in Web
  Applications with Retroactive Auditing}. In \bibinfo{booktitle}{\emph{11th
  {USENIX} Symposium on Operating Systems Design and Implementation ({OSDI}
  14)}}. \bibinfo{publisher}{{USENIX} Association}, \bibinfo{pages}{555--569}.
\newblock
\urldef\tempurl%
\url{https://www.usenix.org/conference/osdi14/technical-sessions/presentation/chen_haogang}
\showURL{%
\tempurl}


\bibitem[\protect\citeauthoryear{Crameri, Knezevic, Kostic, Bianchini, and
  Zwaenepoel}{Crameri et~al\mbox{.}}{2007}]%
        {crameri2007staged}
\bibfield{author}{\bibinfo{person}{Olivier Crameri}, \bibinfo{person}{Nikola
  Knezevic}, \bibinfo{person}{Dejan Kostic}, \bibinfo{person}{Ricardo
  Bianchini}, {and} \bibinfo{person}{Willy Zwaenepoel}.}
  \bibinfo{year}{2007}\natexlab{}.
\newblock \showarticletitle{Staged Deployment in Mirage, an Integrated Software
  Upgrade Testing and Distribution System}.
\newblock \bibinfo{journal}{\emph{ACM SIGOPS Operating Systems Review}}
  \bibinfo{volume}{41}, \bibinfo{number}{6} (\bibinfo{year}{2007}),
  \bibinfo{pages}{221--236}.
\newblock
\urldef\tempurl%
\url{https://doi.org/10.1145/1323293.1294283}
\showDOI{\tempurl}


\bibitem[\protect\citeauthoryear{Dey, Lahiri, and Zhang}{Dey
  et~al\mbox{.}}{2015}]%
        {dey2015optimal}
\bibfield{author}{\bibinfo{person}{Debabrata Dey}, \bibinfo{person}{Atanu
  Lahiri}, {and} \bibinfo{person}{Guoying Zhang}.}
  \bibinfo{year}{2015}\natexlab{}.
\newblock \showarticletitle{Optimal Policies for Security Patch Management}.
\newblock \bibinfo{journal}{\emph{INFORMS Journal on Computing}}
  \bibinfo{volume}{27}, \bibinfo{number}{3} (\bibinfo{year}{2015}),
  \bibinfo{pages}{462--477}.
\newblock
\urldef\tempurl%
\url{https://doi.org/10.1287/ijoc.2014.0638}
\showURL{%
\tempurl}


\bibitem[\protect\citeauthoryear{Dietrich, Krombholz, Borgolte, and
  Fiebig}{Dietrich et~al\mbox{.}}{2018}]%
        {dietrich2018investigating}
\bibfield{author}{\bibinfo{person}{Constanze Dietrich},
  \bibinfo{person}{Katharina Krombholz}, \bibinfo{person}{Kevin Borgolte},
  {and} \bibinfo{person}{Tobias Fiebig}.} \bibinfo{year}{2018}\natexlab{}.
\newblock \showarticletitle{Investigating System Operators’ Perspective on
  Security Misconfigurations}. In \bibinfo{booktitle}{\emph{Proceedings of the
  2018 ACM SIGSAC Conference on Computer and Communications Security
  (CCS’18)}}. \bibinfo{publisher}{ACM}, \bibinfo{pages}{1272--1289}.
\newblock
\urldef\tempurl%
\url{https://doi.org/10.1145/3243734.3243794}
\showURL{%
\tempurl}


\bibitem[\protect\citeauthoryear{Dissanayake, Jayatilaka, Zahedi, and
  Babar}{Dissanayake et~al\mbox{.}}{2021a}]%
        {dissanayake2021software}
\bibfield{author}{\bibinfo{person}{Nesara Dissanayake}, \bibinfo{person}{Asangi
  Jayatilaka}, \bibinfo{person}{Mansooreh Zahedi}, {and}
  \bibinfo{person}{Muhammad~Ali Babar}.} \bibinfo{year}{2021}\natexlab{a}.
\newblock \showarticletitle{Software security patch management-A systematic
  literature review of challenges, approaches, tools and practices}.
\newblock \bibinfo{journal}{\emph{Information and Software Technology}}
  \bibinfo{volume}{144} (\bibinfo{year}{2021}), \bibinfo{pages}{106771}.
\newblock
\urldef\tempurl%
\url{https://doi.org/10.1016/j.infsof.2021.106771}
\showURL{%
\tempurl}


\bibitem[\protect\citeauthoryear{Dissanayake, Zahedi, Jayatilaka, and
  Babar}{Dissanayake et~al\mbox{.}}{2021b}]%
        {dissanayake2021grounded}
\bibfield{author}{\bibinfo{person}{Nesara Dissanayake},
  \bibinfo{person}{Mansooreh Zahedi}, \bibinfo{person}{Asangi Jayatilaka},
  {and} \bibinfo{person}{Muhammad~Ali Babar}.}
  \bibinfo{year}{2021}\natexlab{b}.
\newblock \showarticletitle{A Grounded Theory of the Role of Coordination in
  Software Security Patch Management}. In \bibinfo{booktitle}{\emph{Proceedings
  of the 29th ACM Joint European Software Engineering Conference and Symposium
  on the Foundations of Software Engineering (ESEC/FSE ’21)}}.
  \bibinfo{publisher}{ACM, New York, NY, USA}.
\newblock
\urldef\tempurl%
\url{https://doi.org/10.1145/3468264.3468595}
\showDOI{\tempurl}


\bibitem[\protect\citeauthoryear{Dumitra{\c{s}} and Narasimhan}{Dumitra{\c{s}}
  and Narasimhan}{2009}]%
        {dumitracs2009upgrades}
\bibfield{author}{\bibinfo{person}{Tudor Dumitra{\c{s}}} {and}
  \bibinfo{person}{Priya Narasimhan}.} \bibinfo{year}{2009}\natexlab{}.
\newblock \showarticletitle{Why Do Upgrades Fail and What Can We Do about It?}.
  In \bibinfo{booktitle}{\emph{ACM/IFIP/USENIX International Conference on
  Distributed Systems Platforms and Open Distributed Processing}}
  \emph{(\bibinfo{series}{Middleware 2009. Lecture Notes in Computer Science},
  Vol.~\bibinfo{volume}{5896})}. \bibinfo{publisher}{Springer},
  \bibinfo{address}{Berlin}, \bibinfo{pages}{349--372}.
\newblock
\urldef\tempurl%
\url{https://doi.org/10.1007/978-3-642-10445-9_18}
\showDOI{\tempurl}


\bibitem[\protect\citeauthoryear{Dunagan, Roussev, Daniels, Johnson, Verbowski,
  and Wang}{Dunagan et~al\mbox{.}}{2004}]%
        {dunagan2004towards}
\bibfield{author}{\bibinfo{person}{John Dunagan}, \bibinfo{person}{Roussi
  Roussev}, \bibinfo{person}{Brad Daniels}, \bibinfo{person}{Aaron Johnson},
  \bibinfo{person}{Chad Verbowski}, {and} \bibinfo{person}{Yi-Min Wang}.}
  \bibinfo{year}{2004}\natexlab{}.
\newblock \showarticletitle{Towards a self-managing software patching process
  using black-box persistent-state manifests}. In
  \bibinfo{booktitle}{\emph{International Conference on Autonomic Computing,
  2004. Proceedings.}} \bibinfo{publisher}{IEEE}, \bibinfo{pages}{106--113}.
\newblock
\urldef\tempurl%
\url{https://doi.org/10.1109/ICAC.2004.1301353}
\showDOI{\tempurl}


\bibitem[\protect\citeauthoryear{Eddy and Perlroth}{Eddy and Perlroth}{2020}]%
        {Germancyberattack}
\bibfield{author}{\bibinfo{person}{Melissa Eddy} {and} \bibinfo{person}{Nicole
  Perlroth}.} \bibinfo{year}{2020}\natexlab{}.
\newblock \bibinfo{booktitle}{\emph{Cyber Attack Suspected in German Woman’s
  Death}}.
\newblock
\urldef\tempurl%
\url{https://www.nytimes.com/2020/09/18/world/europe/cyber-attack-germany-ransomeware-death.html?smid=tw-share}
\showURL{%
Retrieved June 23, 2021 from \tempurl}


\bibitem[\protect\citeauthoryear{Genuchten}{Genuchten}{1991}]%
        {van1991software}
\bibfield{author}{\bibinfo{person}{Michiel~Van Genuchten}.}
  \bibinfo{year}{1991}\natexlab{}.
\newblock \showarticletitle{Why is Software Late? An Empirical Study of Reasons
  for Delay in Software Development}.
\newblock \bibinfo{journal}{\emph{IEEE Transactions on Software Engineering}}
  \bibinfo{volume}{17}, \bibinfo{number}{6} (\bibinfo{year}{1991}),
  \bibinfo{pages}{582--590}.
\newblock


\bibitem[\protect\citeauthoryear{Glaser}{Glaser}{1978}]%
        {glaser1978theoretical}
\bibfield{author}{\bibinfo{person}{Barney~G. Glaser}.}
  \bibinfo{year}{1978}\natexlab{}.
\newblock \bibinfo{booktitle}{\emph{Theoretical Sensitivity: Advances in the
  Methodology of Grounded Theory}}.
\newblock \bibinfo{publisher}{Sociology Press}, \bibinfo{address}{Mill Valley,
  CA}.
\newblock


\bibitem[\protect\citeauthoryear{Glaser and Strauss}{Glaser and
  Strauss}{1967}]%
        {glaser1967discovery}
\bibfield{author}{\bibinfo{person}{Barney~G. Glaser} {and}
  \bibinfo{person}{Anselmo~L. Strauss}.} \bibinfo{year}{1967}\natexlab{}.
\newblock \bibinfo{booktitle}{\emph{The Discovery of Grounded Theory:
  Strategies for Qualitative Research}}.
\newblock \bibinfo{publisher}{Aldine Transaction}, \bibinfo{address}{Chicago}.
\newblock


\bibitem[\protect\citeauthoryear{Goodin}{Goodin}{2017}]%
        {Equifaxbug}
\bibfield{author}{\bibinfo{person}{Dan Goodin}.}
  \bibinfo{year}{2017}\natexlab{}.
\newblock \bibinfo{booktitle}{\emph{Failure to patch two-month-old bug led to
  massive Equifax breach}}.
\newblock
\urldef\tempurl%
\url{https://arstechnica.com/information-technology/2017/09/massive-equifax-breach-caused-by-failure-to-patch-two-month-old-bug/}
\showURL{%
Retrieved June 23, 2021 from \tempurl}


\bibitem[\protect\citeauthoryear{Herbsleb and Mockus}{Herbsleb and
  Mockus}{2003}]%
        {1205177}
\bibfield{author}{\bibinfo{person}{James~D. Herbsleb} {and}
  \bibinfo{person}{Audris Mockus}.} \bibinfo{year}{2003}\natexlab{}.
\newblock \showarticletitle{An Empirical Study of Speed and Communication in
  Globally Distributed Software Development}.
\newblock \bibinfo{journal}{\emph{IEEE Transactions on Software Engineering}}
  \bibinfo{volume}{29}, \bibinfo{number}{6} (\bibinfo{year}{2003}),
  \bibinfo{pages}{481--494}.
\newblock
\urldef\tempurl%
\url{https://doi.org/10.1109/TSE.2003.1205177}
\showDOI{\tempurl}


\bibitem[\protect\citeauthoryear{Herbsleb, Mockus, Finholt, and
  Grinter}{Herbsleb et~al\mbox{.}}{2000}]%
        {10.1145/358916.359003}
\bibfield{author}{\bibinfo{person}{James~D. Herbsleb}, \bibinfo{person}{Audris
  Mockus}, \bibinfo{person}{Thomas~A. Finholt}, {and}
  \bibinfo{person}{Rebecca~E. Grinter}.} \bibinfo{year}{2000}\natexlab{}.
\newblock \showarticletitle{Distance, dependencies, and delay in a global
  collaboration}. In \bibinfo{booktitle}{\emph{Proceedings of the 2000 ACM
  Conference on Computer Supported Cooperative Work (CSCW)}}.
  \bibinfo{publisher}{Association for Computing Machinery},
  \bibinfo{pages}{319–328}.
\newblock
\urldef\tempurl%
\url{https://doi.org/10.1145/358916.359003}
\showDOI{\tempurl}


\bibitem[\protect\citeauthoryear{Herbsleb, Mockus, Finholt, and
  Grinter}{Herbsleb et~al\mbox{.}}{2001}]%
        {919083}
\bibfield{author}{\bibinfo{person}{James~D. Herbsleb}, \bibinfo{person}{Audris
  Mockus}, \bibinfo{person}{Thomas~A. Finholt}, {and}
  \bibinfo{person}{Rebecca~E. Grinter}.} \bibinfo{year}{2001}\natexlab{}.
\newblock \showarticletitle{An Empirical Study of Global Software Development:
  Distance and Speed}. In \bibinfo{booktitle}{\emph{Proceedings of the 23rd
  International Conference on Software Engineering, ICSE 2001}}.
  \bibinfo{publisher}{IEEE}, \bibinfo{pages}{81--90}.
\newblock
\urldef\tempurl%
\url{https://doi.org/10.1109/ICSE.2001.919083}
\showDOI{\tempurl}


\bibitem[\protect\citeauthoryear{Huang, Baset, Tang, Gupta, Sudhan, Feroze,
  Garg, and Ravichandran}{Huang et~al\mbox{.}}{2012}]%
        {huang2012patch}
\bibfield{author}{\bibinfo{person}{Hai Huang}, \bibinfo{person}{Salman Baset},
  \bibinfo{person}{Chunqiang Tang}, \bibinfo{person}{Ashu Gupta},
  \bibinfo{person}{KN~Madhu Sudhan}, \bibinfo{person}{Fazal Feroze},
  \bibinfo{person}{Rajesh Garg}, {and} \bibinfo{person}{Sumithra
  Ravichandran}.} \bibinfo{year}{2012}\natexlab{}.
\newblock \showarticletitle{Patch Management Automation for Enterprise Cloud}.
  In \bibinfo{booktitle}{\emph{IEEE Network Operations and Management
  Symposium}}. \bibinfo{publisher}{IEEE}, \bibinfo{pages}{691--705}.
\newblock
\urldef\tempurl%
\url{https://doi.org/10.1109/NOMS.2012.6211988}
\showDOI{\tempurl}


\bibitem[\protect\citeauthoryear{Jenkins, Kalligeros, Vaniea, and
  Wolters}{Jenkins et~al\mbox{.}}{2020}]%
        {jenkins2020anyone}
\bibfield{author}{\bibinfo{person}{Adam Jenkins}, \bibinfo{person}{Pieris
  Kalligeros}, \bibinfo{person}{Kami Vaniea}, {and} \bibinfo{person}{Maria~K.
  Wolters}.} \bibinfo{year}{2020}\natexlab{}.
\newblock \showarticletitle{“Anyone Else Seeing this Error?”: Community,
  System Administrators, and Patch Information}. In
  \bibinfo{booktitle}{\emph{2020 IEEE European Symposium on Security and
  Privacy (EuroS\&P)}}. \bibinfo{publisher}{IEEE}, \bibinfo{pages}{105--119}.
\newblock
\urldef\tempurl%
\url{https://doi.org/10.1109/EuroSP48549.2020.00015}
\showDOI{\tempurl}


\bibitem[\protect\citeauthoryear{Kamar}{Kamar}{2016}]%
        {kamar2016directions}
\bibfield{author}{\bibinfo{person}{Ece Kamar}.}
  \bibinfo{year}{2016}\natexlab{}.
\newblock \showarticletitle{Directions in Hybrid Intelligence: Complementing AI
  Systems with Human Intelligence}. In \bibinfo{booktitle}{\emph{IJCAI}}.
  \bibinfo{publisher}{IEEE}, \bibinfo{pages}{4070--4073}.
\newblock


\bibitem[\protect\citeauthoryear{Kocher, Horn, Fogh, Genkin, Gruss, Haas,
  Hamburg, Lipp, Mangard, Prescher, Schwarz, and Yarom}{Kocher
  et~al\mbox{.}}{2019}]%
        {kocher2019spectre}
\bibfield{author}{\bibinfo{person}{Paul Kocher}, \bibinfo{person}{Jann Horn},
  \bibinfo{person}{Anders Fogh}, \bibinfo{person}{Daniel Genkin},
  \bibinfo{person}{Daniel Gruss}, \bibinfo{person}{Werner Haas},
  \bibinfo{person}{Mike Hamburg}, \bibinfo{person}{Moritz Lipp},
  \bibinfo{person}{Stefan Mangard}, \bibinfo{person}{Thomas Prescher},
  \bibinfo{person}{Michael Schwarz}, {and} \bibinfo{person}{Yuval Yarom}.}
  \bibinfo{year}{2019}\natexlab{}.
\newblock \showarticletitle{Spectre Attacks: Exploiting Speculative Execution}.
  In \bibinfo{booktitle}{\emph{2019 IEEE Symposium on Security and Privacy
  (S\&P’19)}}. \bibinfo{publisher}{IEEE}, \bibinfo{pages}{1--19}.
\newblock
\urldef\tempurl%
\url{https://doi.org/10.1109/SP.2019.00002}
\showDOI{\tempurl}


\bibitem[\protect\citeauthoryear{Li, Rogers, Mathur, Malkin, and Chetty}{Li
  et~al\mbox{.}}{2019}]%
        {li2019keepers}
\bibfield{author}{\bibinfo{person}{Frank Li}, \bibinfo{person}{Lisa Rogers},
  \bibinfo{person}{Arunesh Mathur}, \bibinfo{person}{Nathan Malkin}, {and}
  \bibinfo{person}{Marshini Chetty}.} \bibinfo{year}{2019}\natexlab{}.
\newblock \showarticletitle{Keepers of the Machines: Examining How System
  Administrators Manage Software Updates}. In
  \bibinfo{booktitle}{\emph{Fifteenth Symposium on Usable Privacy and Security
  ({SOUPS} 2019)}}. \bibinfo{publisher}{USENIX Association},
  \bibinfo{pages}{273--288}.
\newblock


\bibitem[\protect\citeauthoryear{Lin, Zagalsky, Storey, and Serebrenik}{Lin
  et~al\mbox{.}}{2016}]%
        {lin2016developers}
\bibfield{author}{\bibinfo{person}{Bin Lin}, \bibinfo{person}{Alexey Zagalsky},
  \bibinfo{person}{Margaret-Anne Storey}, {and} \bibinfo{person}{Alexander
  Serebrenik}.} \bibinfo{year}{2016}\natexlab{}.
\newblock \showarticletitle{Why Developers Are Slacking Off: Understanding How
  Software Teams Use Slack}. In \bibinfo{booktitle}{\emph{Proceedings of the
  19th ACM Conference on Computer Supported Cooperative Work and Social
  Computing Companion (CSCW ’16 Companion)}}. \bibinfo{publisher}{ACM, New
  York, NY, USA}, \bibinfo{pages}{333–336}.
\newblock
\urldef\tempurl%
\url{https://doi.org/10.1145/2818052.2869117}
\showDOI{\tempurl}


\bibitem[\protect\citeauthoryear{Lipp, Schwarz, Gruss, Prescher, Haas, Fogh,
  Horn, Mangard, Kocher, Genkin, Yarom, and Hamburg}{Lipp
  et~al\mbox{.}}{2018}]%
        {217478}
\bibfield{author}{\bibinfo{person}{Moritz Lipp}, \bibinfo{person}{Michael
  Schwarz}, \bibinfo{person}{Daniel Gruss}, \bibinfo{person}{Thomas Prescher},
  \bibinfo{person}{Werner Haas}, \bibinfo{person}{Anders Fogh},
  \bibinfo{person}{Jann Horn}, \bibinfo{person}{Stefan Mangard},
  \bibinfo{person}{Paul Kocher}, \bibinfo{person}{Daniel Genkin},
  \bibinfo{person}{Yuval Yarom}, {and} \bibinfo{person}{Mike Hamburg}.}
  \bibinfo{year}{2018}\natexlab{}.
\newblock \showarticletitle{Meltdown: Reading Kernel Memory from User Space}.
  In \bibinfo{booktitle}{\emph{27th {USENIX} Security Symposium ({USENIX}
  Security 18)}}. \bibinfo{publisher}{{USENIX} Association},
  \bibinfo{pages}{973--990}.
\newblock
\urldef\tempurl%
\url{https://www.usenix.org/conference/usenixsecurity18/presentation/lipp}
\showURL{%
\tempurl}


\bibitem[\protect\citeauthoryear{Marshall}{Marshall}{1996}]%
        {marshall1996sampling}
\bibfield{author}{\bibinfo{person}{Martin~N. Marshall}.}
  \bibinfo{year}{1996}\natexlab{}.
\newblock \showarticletitle{Sampling for qualitative research}.
\newblock \bibinfo{journal}{\emph{Family practice}} \bibinfo{volume}{13},
  \bibinfo{number}{6} (\bibinfo{year}{1996}), \bibinfo{pages}{522--526}.
\newblock
\urldef\tempurl%
\url{https://doi.org/10.1093/fampra/13.6.522}
\showURL{%
\tempurl}


\bibitem[\protect\citeauthoryear{Mathieu, Marks, and Zaccaro}{Mathieu
  et~al\mbox{.}}{2001}]%
        {mathieu2001multi}
\bibfield{author}{\bibinfo{person}{John~E. Mathieu},
  \bibinfo{person}{Michelle~A. Marks}, {and} \bibinfo{person}{Stephen~J.
  Zaccaro}.} \bibinfo{year}{2001}\natexlab{}.
\newblock \showarticletitle{Multiteam systems}.
\newblock \bibinfo{journal}{\emph{Handbook of Industrial, Work and
  Organizational Psychology}}  \bibinfo{volume}{2} (\bibinfo{year}{2001}),
  \bibinfo{pages}{289--313}.
\newblock


\bibitem[\protect\citeauthoryear{Maurer and Brumley}{Maurer and
  Brumley}{2012}]%
        {maurer2012tachyon}
\bibfield{author}{\bibinfo{person}{Matthew Maurer} {and} \bibinfo{person}{David
  Brumley}.} \bibinfo{year}{2012}\natexlab{}.
\newblock \showarticletitle{TACHYON: Tandem execution for efficient live patch
  testing}. In \bibinfo{booktitle}{\emph{21st $\{$USENIX$\}$ Security Symposium
  ($\{$USENIX$\}$ Security 12}}. \bibinfo{publisher}{$\{$USENIX$\}$
  Association}, \bibinfo{address}{Bellevue, WA}, \bibinfo{pages}{617--630}.
\newblock
\urldef\tempurl%
\url{https://www.usenix.org/conference/usenixsecurity12/technical-sessions/presentation/maurer}
\showURL{%
\tempurl}


\bibitem[\protect\citeauthoryear{Maxwell}{Maxwell}{1992}]%
        {maxwell1992understanding}
\bibfield{author}{\bibinfo{person}{Joseph~A. Maxwell}.}
  \bibinfo{year}{1992}\natexlab{}.
\newblock \showarticletitle{Understanding and Validity in Qualitative
  Research}.
\newblock \bibinfo{journal}{\emph{Harvard Educational Review}}
  \bibinfo{volume}{62}, \bibinfo{number}{3} (\bibinfo{year}{1992}),
  \bibinfo{pages}{279--301}.
\newblock
\urldef\tempurl%
\url{https://doi.org/10.17763/haer.62.3.8323320856251826}
\showURL{%
\tempurl}


\bibitem[\protect\citeauthoryear{Mell, Bergeron, and Henning}{Mell
  et~al\mbox{.}}{2005a}]%
        {mell2005creating}
\bibfield{author}{\bibinfo{person}{Peter Mell}, \bibinfo{person}{Tiffany
  Bergeron}, {and} \bibinfo{person}{David Henning}.}
  \bibinfo{year}{2005}\natexlab{a}.
\newblock \showarticletitle{Creating a patch and vulnerability management
  program}.
\newblock \bibinfo{journal}{\emph{NIST Special Publication}}
  \bibinfo{volume}{800} (\bibinfo{year}{2005}), \bibinfo{pages}{40}.
\newblock


\bibitem[\protect\citeauthoryear{Mell, Bergeron, and Henning}{Mell
  et~al\mbox{.}}{2005b}]%
        {nist2005guide}
\bibfield{author}{\bibinfo{person}{Peter Mell}, \bibinfo{person}{Tiffany
  Bergeron}, {and} \bibinfo{person}{David Henning}.}
  \bibinfo{year}{2005}\natexlab{b}.
\newblock \showarticletitle{Creating a Patch and Vulnerability Management
  Program}.
\newblock \bibinfo{journal}{\emph{NIST Special Publication (SP) 800-40 Revision
  2}} (\bibinfo{year}{2005}).
\newblock


\bibitem[\protect\citeauthoryear{Merriam}{Merriam}{1998}]%
        {merriam1998qualitative}
\bibfield{author}{\bibinfo{person}{Sharan~B. Merriam}.}
  \bibinfo{year}{1998}\natexlab{}.
\newblock \bibinfo{booktitle}{\emph{Qualitative Research and Case Study
  Applications in Education. Revised and Expanded from}}.
\newblock \bibinfo{publisher}{ERIC}.
\newblock


\bibitem[\protect\citeauthoryear{Nappa, Johnson, Bilge, Caballero, and
  Dumitras}{Nappa et~al\mbox{.}}{2015}]%
        {nappa2015attack}
\bibfield{author}{\bibinfo{person}{Antonio Nappa}, \bibinfo{person}{Richard
  Johnson}, \bibinfo{person}{Leyla Bilge}, \bibinfo{person}{Juan Caballero},
  {and} \bibinfo{person}{Tudor Dumitras}.} \bibinfo{year}{2015}\natexlab{}.
\newblock \showarticletitle{The Attack of the Clones: A Study of the Impact of
  Shared Code on Vulnerability Patching}. In \bibinfo{booktitle}{\emph{IEEE
  Symposium on Security and Privacy (S\&P)}}. \bibinfo{publisher}{IEEE},
  \bibinfo{pages}{692--708}.
\newblock
\urldef\tempurl%
\url{https://doi.org/10.1109/SP.2015.48}
\showDOI{\tempurl}


\bibitem[\protect\citeauthoryear{Newman}{Newman}{2017}]%
        {Equifaxwired}
\bibfield{author}{\bibinfo{person}{Lily~Hay Newman}.}
  \bibinfo{year}{2017}\natexlab{}.
\newblock \bibinfo{booktitle}{\emph{Equifax Officially Has No Excuse}}.
\newblock
\urldef\tempurl%
\url{https://www.wired.com/story/equifax-breach-no-excuse/}
\showURL{%
Retrieved June 23, 2021 from \tempurl}


\bibitem[\protect\citeauthoryear{Nguyen, Wolf, and Damian}{Nguyen
  et~al\mbox{.}}{2008}]%
        {nguyen2008global}
\bibfield{author}{\bibinfo{person}{Thanh Nguyen}, \bibinfo{person}{Timo Wolf},
  {and} \bibinfo{person}{Daniela Damian}.} \bibinfo{year}{2008}\natexlab{}.
\newblock \showarticletitle{Global Software Development and Delay: Does
  Distance Still Matter?}. In \bibinfo{booktitle}{\emph{2008 IEEE International
  Conference on Global Software Engineering}}. \bibinfo{publisher}{IEEE},
  \bibinfo{pages}{45--54}.
\newblock
\urldef\tempurl%
\url{https://doi.org/10.1109/ICGSE.2008.39}
\showDOI{\tempurl}


\bibitem[\protect\citeauthoryear{Nicastro}{Nicastro}{2003}]%
        {nicastro2003security}
\bibfield{author}{\bibinfo{person}{Felicia~M. Nicastro}.}
  \bibinfo{year}{2003}\natexlab{}.
\newblock \showarticletitle{Security Patch Management}.
\newblock \bibinfo{journal}{\emph{Inf. Secur. J. A Glob. Perspect.}}
  \bibinfo{volume}{12}, \bibinfo{number}{5} (\bibinfo{year}{2003}),
  \bibinfo{pages}{5--18}.
\newblock
\urldef\tempurl%
\url{https://doi.org/10.1201/1086/43808.12.5.20031101/78486.2}
\showURL{%
\tempurl}


\bibitem[\protect\citeauthoryear{NIST}{NIST}{2002}]%
        {nist2002guide}
\bibfield{author}{\bibinfo{person}{NIST}.} \bibinfo{year}{2002}\natexlab{}.
\newblock \showarticletitle{Procedures for Handling Security Patches}.
\newblock \bibinfo{journal}{\emph{Special Publication (SP) 800-40}}
  (\bibinfo{year}{2002}).
\newblock


\bibitem[\protect\citeauthoryear{Post and Kagan}{Post and Kagan}{2003}]%
        {post2003computer}
\bibfield{author}{\bibinfo{person}{Gerald Post} {and} \bibinfo{person}{Albert
  Kagan}.} \bibinfo{year}{2003}\natexlab{}.
\newblock \showarticletitle{Computer security and operating system updates}.
\newblock \bibinfo{journal}{\emph{Information and Software Technology}}
  \bibinfo{volume}{45}, \bibinfo{number}{8} (\bibinfo{year}{2003}),
  \bibinfo{pages}{461--467}.
\newblock


\bibitem[\protect\citeauthoryear{Potter and Nieh}{Potter and Nieh}{2005}]%
        {potter2005reducing}
\bibfield{author}{\bibinfo{person}{Shaya Potter} {and} \bibinfo{person}{Jason
  Nieh}.} \bibinfo{year}{2005}\natexlab{}.
\newblock \showarticletitle{Reducing Downtime Due to System Maintenance and
  Upgrades}. In \bibinfo{booktitle}{\emph{Proceedings of the 19th USENIX
  Systems Administration Conference}}. \bibinfo{publisher}{IEEE},
  \bibinfo{pages}{6--6}.
\newblock


\bibitem[\protect\citeauthoryear{Rodriguez, Urquhart, and Mendes}{Rodriguez
  et~al\mbox{.}}{2020}]%
        {rodriguez2020theory}
\bibfield{author}{\bibinfo{person}{Pilar Rodriguez}, \bibinfo{person}{Cathy
  Urquhart}, {and} \bibinfo{person}{Emilia Mendes}.}
  \bibinfo{year}{2020}\natexlab{}.
\newblock \showarticletitle{A Theory of Value for Value-based Feature Selection
  in Software Engineering}.
\newblock \bibinfo{journal}{\emph{IEEE Transactions on Software Engineering}}
  (\bibinfo{year}{2020}).
\newblock
\urldef\tempurl%
\url{https://doi.org/10.1109/TSE.2020.2989666}
\showDOI{\tempurl}


\bibitem[\protect\citeauthoryear{Runeson and H{\"o}st}{Runeson and
  H{\"o}st}{2009}]%
        {runeson2009guidelines}
\bibfield{author}{\bibinfo{person}{Per Runeson} {and} \bibinfo{person}{Martin
  H{\"o}st}.} \bibinfo{year}{2009}\natexlab{}.
\newblock \showarticletitle{Guidelines for conducting and reporting case study
  research in software engineering}.
\newblock \bibinfo{journal}{\emph{Empirical Software Engineering}}
  \bibinfo{volume}{14}, \bibinfo{number}{2} (\bibinfo{year}{2009}),
  \bibinfo{pages}{131--164}.
\newblock


\bibitem[\protect\citeauthoryear{Schwandt}{Schwandt}{1997}]%
        {schwandt1997sage}
\bibfield{author}{\bibinfo{person}{Thomas~A. Schwandt}.}
  \bibinfo{year}{1997}\natexlab{}.
\newblock \bibinfo{booktitle}{\emph{Qualitative Inquiry}}.
\newblock \bibinfo{publisher}{Sage}, \bibinfo{address}{London}.
\newblock


\bibitem[\protect\citeauthoryear{Security}{Security}{2020}]%
        {Accenturereport2020}
\bibfield{author}{\bibinfo{person}{Accenture Security}.}
  \bibinfo{year}{2020}\natexlab{}.
\newblock \bibinfo{booktitle}{\emph{2020 Cyber Threatscape Report}}.
\newblock
\urldef\tempurl%
\url{https://www.accenture.com/_acnmedia/PDF-136/Accenture-2020-Cyber-Threatscape-Full-Report.pdf}
\showURL{%
Retrieved June 23, 2021 from \tempurl}


\bibitem[\protect\citeauthoryear{Souppaya and Scarfone}{Souppaya and
  Scarfone}{2013}]%
        {souppaya2013guide}
\bibfield{author}{\bibinfo{person}{Murugiah Souppaya} {and}
  \bibinfo{person}{Karen Scarfone}.} \bibinfo{year}{2013}\natexlab{}.
\newblock \showarticletitle{Guide to Enterprise Patch Management Technologies}.
\newblock \bibinfo{journal}{\emph{NIST Special Publication 800-40 Revision 3}}
  (\bibinfo{year}{2013}), \bibinfo{pages}{40}.
\newblock
\urldef\tempurl%
\url{http://dx.doi.org/10.6028/NIST.SP.800-40r3}
\showURL{%
\tempurl}


\bibitem[\protect\citeauthoryear{Strauss and Corbin}{Strauss and
  Corbin}{1998}]%
        {strauss1998basics}
\bibfield{author}{\bibinfo{person}{Anselm~L. Strauss} {and}
  \bibinfo{person}{Juliet~M. Corbin}.} \bibinfo{year}{1998}\natexlab{}.
\newblock \bibinfo{booktitle}{\emph{Basics of Qualitative Research : Techniques
  and Procedures for Developing Grounded Theory} (\bibinfo{edition}{2nd} ed.)}.
\newblock \bibinfo{publisher}{Sage}.
\newblock


\bibitem[\protect\citeauthoryear{Strauss and Corbin}{Strauss and
  Corbin}{2007}]%
        {strauss2007basics}
\bibfield{author}{\bibinfo{person}{Anselm~L. Strauss} {and}
  \bibinfo{person}{Juliet~M. Corbin}.} \bibinfo{year}{2007}\natexlab{}.
\newblock \bibinfo{booktitle}{\emph{Basics of Qualitative Research: Techniques
  and Procedures for Developing Grounded Theory} (\bibinfo{edition}{3rd} ed.)}.
\newblock \bibinfo{publisher}{Sage}.
\newblock


\bibitem[\protect\citeauthoryear{Tiefenau, Häring, Krombholz, and von
  Zezschwitz}{Tiefenau et~al\mbox{.}}{2020}]%
        {tiefenau2020security}
\bibfield{author}{\bibinfo{person}{Christian Tiefenau},
  \bibinfo{person}{Maximilian Häring}, \bibinfo{person}{Katharina Krombholz},
  {and} \bibinfo{person}{Emanuel von Zezschwitz}.}
  \bibinfo{year}{2020}\natexlab{}.
\newblock \showarticletitle{Security, Availability, and Multiple Information
  Sources: Exploring Update Behavior of System Administrators}. In
  \bibinfo{booktitle}{\emph{Sixteenth Symposium on Usable Privacy and Security
  (SOUPS 2020)}}. \bibinfo{publisher}{USENIX Association},
  \bibinfo{pages}{239--258}.
\newblock


\bibitem[\protect\citeauthoryear{Tucek, Xiong, and Zhou}{Tucek
  et~al\mbox{.}}{2009}]%
        {tucek2009efficient}
\bibfield{author}{\bibinfo{person}{Joseph Tucek}, \bibinfo{person}{Weiwei
  Xiong}, {and} \bibinfo{person}{Yuanyuan Zhou}.}
  \bibinfo{year}{2009}\natexlab{}.
\newblock \showarticletitle{Efficient online validation with delta execution}.
  In \bibinfo{booktitle}{\emph{Proceedings of the 14th international conference
  on Architectural support for programming languages and operating systems}}.
  \bibinfo{pages}{193--204}.
\newblock


\bibitem[\protect\citeauthoryear{Urquhart}{Urquhart}{2013}]%
        {urquhart2012grounded}
\bibfield{author}{\bibinfo{person}{Cathy Urquhart}.}
  \bibinfo{year}{2013}\natexlab{}.
\newblock \bibinfo{booktitle}{\emph{Grounded Theory for Qualitative Research: A
  Practical Guide}}.
\newblock \bibinfo{publisher}{Sage}.
\newblock


\bibitem[\protect\citeauthoryear{Wang, Weisz, Muller, Ram, Geyer, Dugan,
  Tausczik, Samulowitz, and Gray}{Wang et~al\mbox{.}}{2019}]%
        {wang2019human}
\bibfield{author}{\bibinfo{person}{Dakuo Wang}, \bibinfo{person}{Justin~D.
  Weisz}, \bibinfo{person}{Michael Muller}, \bibinfo{person}{Parikshit Ram},
  \bibinfo{person}{Werner Geyer}, \bibinfo{person}{Casey Dugan},
  \bibinfo{person}{Yla~R Tausczik}, \bibinfo{person}{Horst Samulowitz}, {and}
  \bibinfo{person}{Alexander Gray}.} \bibinfo{year}{2019}\natexlab{}.
\newblock \showarticletitle{Human-AI Collaboration in Data Science: Exploring
  Data Scientists’ Perceptions of Automated AI}.
\newblock \bibinfo{journal}{\emph{Proceedings of the ACM on Human-Computer
  Interaction}} \bibinfo{volume}{3}, \bibinfo{number}{CSCW}
  (\bibinfo{year}{2019}).
\newblock
\urldef\tempurl%
\url{https://doi.org/10.1145/3359313}
\showDOI{\tempurl}


\bibitem[\protect\citeauthoryear{Wessel, de~Souza, Steinmacher, Wiese, Polato,
  Chaves, and Gerosa}{Wessel et~al\mbox{.}}{2018}]%
        {10.1145/3274451}
\bibfield{author}{\bibinfo{person}{Mairieli Wessel},
  \bibinfo{person}{Bruno~Mendes de Souza}, \bibinfo{person}{Igor Steinmacher},
  \bibinfo{person}{Igor~S. Wiese}, \bibinfo{person}{Ivanilton Polato},
  \bibinfo{person}{Ana~Paula Chaves}, {and} \bibinfo{person}{Marco~A. Gerosa}.}
  \bibinfo{year}{2018}\natexlab{}.
\newblock \showarticletitle{The Power of Bots: Characterizing and Understanding
  Bots in OSS Projects}.
\newblock \bibinfo{journal}{\emph{Proceedings of the ACM Conference on Computer
  Supported Cooperative Work Social Computing}} \bibinfo{volume}{2},
  \bibinfo{number}{CSCW} (\bibinfo{year}{2018}).
\newblock
\urldef\tempurl%
\url{https://doi.org/10.1145/3274451}
\showDOI{\tempurl}


\bibitem[\protect\citeauthoryear{Yin}{Yin}{1994}]%
        {yin1994case}
\bibfield{author}{\bibinfo{person}{Robert~K. Yin}.}
  \bibinfo{year}{1994}\natexlab{}.
\newblock \bibinfo{title}{Case Study Research: Design and Methods}.
\newblock
\newblock


\bibitem[\protect\citeauthoryear{Yin}{Yin}{2009}]%
        {yin2009case}
\bibfield{author}{\bibinfo{person}{Robert~K. Yin}.}
  \bibinfo{year}{2009}\natexlab{}.
\newblock \bibinfo{booktitle}{\emph{Case Study Research: Design and Methods}
  (\bibinfo{edition}{4} ed.)}.
\newblock \bibinfo{publisher}{Sage}, \bibinfo{address}{Thousand Oaks, CA, USA}.
\newblock


\end{thebibliography}

% For CSCW2 Article 356-378, use
\received{July 2021}
\received[revised]{November 2021}
\received[accepted]{February 2022}

\appendix

% \section{Example of the Patching Tracker}
% \label{appendix:patchingtrackerexample}

% \begin{figure}[h]
%   \includegraphics[scale=0.25]{Figures/Patching_Tracker_Abstract.png}
%   \caption{An extract from team EMR's Patching Tracker - 19.05.2021.}
%   \Description{An extract from the Patching Tracker.}
%   \label{fig:dataanalysis}
% \end{figure}

\end{document}